\documentclass[twoside,a4paper]{article}
\usepackage[english]{babel}
\usepackage[letterpaper,top=3cm,bottom=3cm,left=3cm,right=3cm,marginparwidth=1.75cm]{geometry}
\usepackage{amsmath,graphicx,amssymb,fancyhdr,amsthm,tensor}
\usepackage{changepage}
\usepackage{accents}
\usepackage[colorlinks=true, allcolors=blue]{hyperref}
\usepackage{subcaption}

\begin{document}

\begin{center}
{\Large \bf Black Hole Persistence in New General Relativity \\ }
\vskip 0.7cm
{B. Yildirim\footnote{Corresponding author: bl541771@dal.ca}, A. A. Coley, and D. F. L\'opez}
\vskip 0.2cm
{\it Department of Mathematics and Statistics\\ 
Dalhousie University\\
Halifax, Canada}\\
\vskip 1cm

\begin{quote}
{\bf Abstract.}~{\small We investigate whether black holes can persist through the bounce with a minimal scale factor in a non-singular cosmology, whereby black holes from a previous contracting phase survive into the current expanding one. We do so by studying a generalized McVittie spacetime which embeds a spherically symmetric  black hole in a positive spatial curvature bouncing FLRW cosmological background within the modified theory of teleparallel new general relativity. There are no further assumptions on the spacetime (e.g., on the form of the scale factor) initially, and the local evolution is derived from the field equations of the theory, utilizing a perturbative scheme which is valid ``near the bounce". To leading order we obtain a simple bounce solution similar to that in general relativity for a closed FLRW model with a positive cosmological constant, but in which the curvature term in the Friedmann equation is re-normalized within new general relativity. Qualitatively the minimum of the bounce at $t=0$  changes, but near the bounce the evolution remains symmetric. The central inhomogeneity evolves at higher perturbative orders, where the details depend on the arbitrary constants of the  perturbative solution. Hence the evolution of the local horizon during the bounce changes qualitatively, where the effects depend on the signs of the perturbation, and the symmetry across the bounce is disrupted due to a linear term.
}
\end{quote}
\end{center}
\vskip 1.0cm

\newpage

\section{Introduction}

There are many cosmological models in which the current expansion of the Universe would have followed a collapsing period. A viable cosmological alternative to the standard $\Lambda$CDM cosmology based on the Friedmann-Lemaître-Robertson-Walker (FLRW) model, which also provides a relativistic-compatible solution to the singularity problem, is based on a non-singular bouncing cosmology \cite{novello}, whereby an initially contracting phase connects with the currently expanding phase through some minimal scale factor (and hence a vanishing Hubble rate). These models were first proposed in \cite{Tolman31,Lemaitre33,Lemaitre:1933gd}. The  bounce which connects these two phases may be generated by classical effects (such as a cosmological constant \cite{lemaitre}, higher-derivative terms \cite{novello}), semi-quantum gravitational effects (associated with string theory \cite{string}, the ``pre-big-bang'' scenario \cite{BHScorr}, loop quantum gravity \cite{sing}), and quantum cosmological effects \cite{ashtekar}. A bounce may also generate the density fluctuations required to make galaxies \cite{quintin} and thereby eliminate one motivation for inflation \cite{Battefeld_2015}. Bouncing cosmologies were reviewed in \cite{brand} (see also \cite{novello} \cite{Battefeld_2015}).  A universe which goes through periods of expansion and compression is called a cyclic universe \cite{cyclic}.

The key to producing a bounce is to break one of the energy conditions (ECs) \cite{Battefeld_2015}. There are two ways to achieve this: the first one operates within GR and requires the violation of the null EC, due to the appearance of fields with negative kinetic energy \cite{Martin:2003sf}. The second allows for a classically singular bounce via an alternative theory of gravity. Nonsingular bouncing cosmologies require non-standard matter terms, or modifications to Einstein's General Relativity (GR) (for example, Horndeski theories). In particular, a classical bounce can occur in models with a scalar field in which the null energy condition (NEC) is violated. For example, Galkina et al \cite{Galkina_2019} investigated regular bouncing cosmologies in Brans-Dicke theory.

In addition to enabling inflationary models, a bounce can be generated by a galileon field \cite{Nicolis_2009Galileon,Creminelli:2010ba,Creminelli:2012my}, often referred to as a \emph{G-bounce} \cite{Qiu:2011cy,Osipov:2013ssa,Easson:2011zy}. These Galileon theories provide a well behaved way to induce a bounce by avoiding ghosts and instabilities \cite{Battefeld_2015}. Ijjas and Steinhardt \cite{ijjas} invoke a scalar field with a cubic Galileon action. They claim that a classically stable non-singular bounce is possible. However, in \cite{lib} it is claimed that singularities are unavoidable in NEC-violating Galileon theories. Likewise, in \cite{Dobre:2017pnt} it is shown that the solution in \cite{ijjas} exists for just 168 Planck times between two singularities, one with an infinite speed of sound and the other with a vanishing speed of sound. Both correspond to infinitely strong coupling in the very same way as a cosmological singularity.

Other scalar-tensor theories have been used to generate a bounce such as the Quintom bounce with its two interacting scalar fields \cite{Feng:2004ad,Feng:2004ff}, in massive gravity \cite{deRham:2010ik,deRham:2010kj,Biswas:2010zk} using the graviton has led to a ghost-free non-singular bounce and cyclic cosmological evolutions at early times \cite{Cai:2012ag}, and with a standard scalar field \cite{Bacalhau_2018}.

String theorists have also studied cosmological solutions in dilaton gravity, a scalar-tensor effective theory for string theory, leading to the pre-big bang (PBB) scenario with a non singular bounce \cite{Gasperini:2002bn}, initiating renewed interest in bouncing scenarios and providing a challenge to the inflationary paradigm \cite{Battefeld_2015}. To study bouncing models with string theory, often a low energy effective theory is used which takes the form of GR with the addition of some scalar field \cite{Battefeld_2015}. The problem with these models is the uncertainty of when effective theories are valid, as well as the appearance of singularities and ghosts \cite{Battefeld_2015}. One of these string theory bounces is called the \emph{Ekpyrotic scenario} \cite{Khoury:2001wf} based on the
5-dimensional heterotic string theory, in which the fifth dimension ends at two different boundary branes whose collision produces a bounce. Another is the {\it cyclic} {\it scenario}, which is an extension of the ekpyrotic scenario by making it cyclic and can be described in terms of an effective 4-dimensional scalar field with a specific potential \cite{Steinhardt:2001st,ijjas,Khoury:2001bz}.

Some less common models used to generate a bounce include: Ho\v{r}ava-Liftshitz (HL) gravity \cite{Horava:2009uw,Horava:2009if} which is a modification to GR at high energies; the possibility of a non-singular bounce was studied in \cite{Kiritsis:2009sh,Brandenberger:2009yt}. $F(R)$ theories ~\cite{DeFelice:2010aj}, constructed to produce a bounce, were studied in \cite{Bamba:2013fha}. A Lagrangian function of the Gauss-Bonnet invariant instead of one of the curvature Ricci scalar $R$ can also be considered; bouncing solutions were explored in \cite{Bamba:2014mya}. An Effective field theory for Quantum Gravity allows spatial derivatives in the action up to 6th order, which then allows for bouncing and cyclic solutions \cite{Cai:2009in}. A less common method of modifying gravity to produce a bounce is provided by the Einstein-Cartan-Sciama-Kibble (ECSK) theory \cite{Kibble:1961ba,Sciama:1964wt}, which uses torsion in addition to curvature as field strengths. In teleparallel theories, we can consider bouncing $F(T)$ theories \cite{Cai:2011tc}, which are similar to $F(R)$ theories but have the advantage that their field equations remain second order which reduces the risk of instabilities. New General Relativity (NGR) \cite{Bahamonde_2023}, perhaps a more conservative teleparallel extension of GR, can also exhibit a bouncing solution, through a very similar mechanism to the one in GR, as we shall see below. Finally, we can also consider a bounce generated by quantum gravity effects. For example, with loop quantum gravity, the most common approach consists in taking a modified Friedmann equation containing a $-\rho^2$ contribution to the right hand side
\cite{Ashtekar:2006rx,WilsonEwing:2012pu}, thus inducing a bounce. There are some interesting additional questions if the bounce is not classical but is quantum in nature. If the bounce occurs due to quantum cosmological effects, then it presumably occurs at the Planck density. In this case, primarily only Planck mass black holes could persist through the bounce.  The possibility of bouncing universes in the context of the no boundary wave function of quantum cosmology \cite{hh} has been considered by Hartle and Hertog \cite{hertog}. 

An important open question is what happens in bouncing cosmologies in the inhomogeneous and non-perturbative regime. There are relatively few studies of the dynamics of the bounce \cite{ijjas}. No black holes can form before the Planck density but they can be expected to form during the contraction of matter and radiation dominated universes \cite{quintin} and will generally be present from previous eras in cyclic cosmologies \cite{cc,ccc}.

\subsection{Persistence}

There is the interesting possibility~\cite{cc} that black holes can persist (``pre-big-bang black holes'' \\ (PBBBHs)) in a Universe which re-collapses to a big crunch and subsequently bounces into a new expansion phase \cite{cc}, so that some black holes from a previous era may have survived the big bounce. There are two possibilities:  (i) ``pre-crunch black holes'' (PCBHs) which ought to have similar characteristics as the black holes which form in the present universe and (ii) black holes which could be generated by the high density at the bounce (``big-crunch black holes'' (BCBHs)) which could be much smaller.  These would be distinct from the ``primordial black holes'' (PBHs) which formed just after the big bang. 

In  \cite{ccc} some 4-dimensional dynamical exact solutions  which describe a regular lattice of black holes within a cosmological background dominated by a scalar field at the bounce were derived, showing that there do indeed exist exact solutions in which multiple black holes persist through a bounce. Since the bounce need not occur at the Planck scale, a classical approach is fully justified.  The previous emphasis has been  mathematical rather than physical, so that it is of interest to examine the cosmological applications in more detail, including whether persistent black holes might be {\em{seeds}} for dark matter and LSS in the current epoch and play a similar role to PBHs in this respect. 

If PBHs or PBBBHs were big enough, they would still be around today and they could be the answer to many cosmological puzzles, from dark matter to supermassive black holes \cite{cc}. The possibility that dark matter could be made up of PBHs has become topical \cite{carr-silk}. There are many constraints on the mass range of such PBHs but there are three mass windows in which this is possible \cite{cks}: around the Planck mass ($10^{-5}$g) if stable relics are left by evaporating black holes; the sublunar mass range ($10^{20} - 10^{24}$g); and the intermediate mass range ($10 -10^3 M_{\odot}$). In various mass ranges, PBBBHs could also contribute to dark matter, provide seeds for galaxies, generate entropy and even drive the bounce itself.

Indeed, Rovelli and Vidotto~\cite{rovelli} have investigated the possibility that dark matter is made up of the remnants of PBBBHs. They argued that only a tiny fraction of the volume of the universe would be outside these black holes at the bounce, but observers in these regions would see a spatially homogeneous universe at later times. This  raises the question of whether the same black holes could make up the dark matter in successive cosmic cycles, with the fraction of dark matter progressively increasing. 
A recent study \cite{cainew} suggests that we might eventually be able to distinguish PBBBHs from black holes formed in our current universe since we know that they already existed very early (possibly too soon for them to have been created by standard astrophysical processes), and it isn’t clear how they could grow so big so fast unless they were seeded. 

In the standard cosmological paradigm, spatial homogeneity is only statistically valid on scales larger than approximately 100-115 Mpc \cite{1}, and there can only exist spatial inhomogeneities up to scales of about 100 Mpc. However, the actual Universe has web-like features which are dominated by huge voids within bubble walls, and
with filamentary structures which contain clusters of galaxies; indeed, the distribution of matter is definitely not spatially homogeneous on scales less than 150-300 Mpc. Within the currently accepted theory of hierarchical structure formation in standard cosmology, it is difficult to understand how galaxies could accumulate so much mass through mergers or accretion alone, and predict that the first galaxies should only appear between redshifts of approximately $15 \leq z \leq 20$ \cite{13}. This raises the question of whether there is sufficient time for these objects to grow to a BH with mass  $\sim 10^9 - 10^{10}M_\odot$ in such a short time after the big bang.

These issues have been exacerbated more recently by observations at the James Webb Space Telescope (JWST) \cite{14}, in which the emergence of a population of (unusually) massive galaxy sources ($> 10^{10} M_\odot$) at $z > 7 - 10$, less than 700 Myr after the Big Bang \cite{17}, have been identified.  Also recent JWST observations have observed a massive BH with MBH $\sim 10^7 - 10^8M_\odot$ at $z \sim 10.3$ \cite{19} and high-redshift quasars, which reveal that many super massive black holes (SMBHs) existed 700 Million years after the big bang or less. Therefore, JWST observations could challenge the standard $\Lambda$CDM cosmology \cite{20}.
 
A variety of possible solutions to the problem of the emergence of SMBHs when the Universe was around $\sim$ 800 Myr have been investigated, which might still reconcile observations with theory \cite{13}.  A popular solution might be via an enhancement of the gravitational force. Some researchers utilize dark matter halos \cite{23}, while other authors prefer a solution by using an alternative theory of gravity to GR \cite{1}. Perhaps more conventionally, massive PBHs or PBBBHs, which were present in the early universe at a pre-stellar epoch, could seed galaxy and quasar formation in the very young universe \cite{carr-silk}.  A population of red candidate massive galaxies ($> 10^{10}M_\odot$) at $7.4 \leq z \leq 9.1$, 500–700 Myr after the big bang have been observed \cite{17}.

\subsection{Recent work}

Recently, Corman et al.~\cite{corman} and P{\`e}rez and  Romero~\cite{perez} have calculated the behaviour of a {\it single} black hole during a bounce. Although the details of their calculations differ, both groups concluded that the black hole could survive through the bounce.

Corman et al.~\cite{corman} numerically studied the evolution of black holes classicaly during a non-singular cosmological bounce utilizing a simple bouncing cosmological model driven by the combination of a ghost scalar field (treated as an effective model for NCC violation) and an ordinary
scalar field, using the nonlinear evolution of the Einstein field equations to follow black holes of various sizes through the bounce. In particular, asymptotically cosmological initial data was tuned to allow for contraction, which is then followed by a bounce that ends with cosmological expansion. 

The violation of the NEC allows for a shrinking black hole horizon and it was shown that for sufficiently large black holes the apparent horizon can even disappear during the contraction phase. It was also shown that independent of the initial mass of the black hole, its event horizon persists throughout the bounce, and that subsequently the late time evolution is that of an expanding universe with a black hole whose mass is comparable to its original value \cite{corman}. This means that black holes created (or already present) in the contraction phase \cite{quintin} can always persist through the bounce and hence have observational consequences in the post-bounce era.

The McVittie spacetime \cite{Mcvittieoriginal} is an analytic solution of the Einstein field equations with perfect fluid source that embeds a spherically symmetric Schwarzschild black hole in a spatially flat FLRW cosmological background. The global properties of the McVittie spacetime were investigated in \cite{A107,A110}. The question as to whether a black hole can persist in a universe that undergoes a cosmological bounce was investigated analytically in \cite{perez}
in a bouncing cosmological background based on a generalization of the spatially flat McVittie spacetime \cite{Faraoni_2007}. 

In \cite{perez}, only the spatially ($k=0$) flat generalized McVittie solutions were considered. The work was also limited by the fact that, since the governing equations are under-determined, the metric evolution is prescribed in an ad-hoc manner, and from that the implied matter (imperfect fluid) type and evolution is derived. In addition, a scale factor ansatz, albeit motivated by quantum theory \cite{Peter_2007,Pinto_Neto_2013}, was chosen [i.e., no scalar (phantom field) was used] to provide a classical bounce (so the model is not a solution [per se]). It was shown \cite{perez} that the horizon associated with this metric changes with cosmic time because it is coupled to the cosmic evolution. The area of the horizon of the black hole decreases during the contracting phase, reaching its smallest value at the bounce. Subsequently, in the expanding phase, its surface area starts to increase proportionally to the square of the scale factor. In the McVittie spacetime, this does not occur, since the horizon of the black hole merges with the cosmological horizon before the bounce is reached.

\section{McVittie solution}

The original McVittie metric \cite{Mcvittieoriginal} attempts to model a black hole in a cosmological background. However, we can consider a generalization of the original McVittie metric by considering

\begin{equation}
    ds^2 =-\left( \frac{1-\frac{m(t)}{2r}\omega}{1+\frac{m(t)}{2r} \omega} \right)^2dt^2 + a(t)^2 \frac{\left( 1+\frac{m(t)}{2r} \omega \right)^4 }{\omega^4}(dr^2 + r^2d\Omega^2) \label{mcvittie}
\end{equation}

\noindent where $\omega \equiv \sqrt{1+ \frac{k}{4}r^2}$, which becomes the standard flat McVittie metric when $m(t) = \frac{m_0}{a(t)}$ and $k=0$ \cite{Mcvittieoriginal}. This choice for the mass function is called the \emph{non-accretion condition} and comes from solving $G \indices{^0 _1} = 0$. This condition also causes the pressure to diverge on the black hole horizon, and so one may assume that the singularity would vanish if we were to allow for such accretion \cite{Faraoni_2007}. Indeed, this is supported by the fact that the divergence of the Ricci scalar is caused by the divergence of the pressure on the horizon \cite{Faraoni_2007}. In addition, the singularity on the horizon is gravitationally weak, which is the case if the volume elements all have finite non-zero limits at the singularity \cite{Nolan_1999}. More recently it has been argued by Kaloper et al \cite{Kaloper_2010} that the introduction of a positive cosmological constant to the flat McVittie solution eliminates the horizon singularity, as then the space-time asymptotes to that of Schwarzchild-de Sitter which resolves the singularity since the singular surface exists at a finite distance but at $t = \infty$.

A global issue with this metric \eqref{mcvittie} is that when $k=\pm 1$, the background cosmological fluid cannot be homogeneous and consequently the metric does not asymptote to any FLRW solution as $r \to \infty$ \cite{Nandra_2012}. To avoid this issue, a similar metric that describes a point mass embedded in an FLRW universe was constructed in \cite{Nandra_2012} and \cite{A107}, where the latter is shown to describe a spatially closed FLRW background cosmology with a central point mass that can describe a black hole. However, we will not be concerned with global behaviour because we will conduct a local analysis near the bounce and hence, the solution we find can be interpreted as an effective theory that describes the behaviour of the cosmology near the bounce.

Perez and Romero \cite{perez} generalized the standard flat McVittie metric by relaxing the non-accretion condition as above. To avoid reduction to the standard McVittie metric once the field equations are solved, this generalization requires a non-zero off-diagonal component in the energy-momentum tensor. The McVittie metric, and its generalizations, model a central black hole in a dynamic cosmological background. This can be seen by taking the limit $m \rightarrow 0$ of \eqref{mcvittie} which gives the isotropic FLRW metric

\begin{equation}
    ds^2 = -dt^2+ \frac{a(t)^2}{(1+\frac{k}{4}r^2)^2} (dr^2 + r^2 d \Omega^2) \label{isoflrw}
\end{equation}

\noindent Also, we can obtain the isotropic Schwarzchild metric if we take the limit $a(t) \rightarrow 1$ and $\omega(r) \rightarrow 1$

\begin{equation}
        ds^2 =-\left( \frac{1-\frac{m}{2r}}{1+\frac{m}{2r} } \right)^2dt^2 + \left( 1+\frac{m}{2r}  \right)^4 (dr^2 + r^2d\Omega^2) \label{isoschw}
\end{equation}

\subsection{Standard McVittie GR Metric}

In GR, the original McVittie metric is often considered (equation \eqref{mcvittie} with $m(t)=\frac{m_0}{a(t)}$ and $k=0$) with the energy-momentum tensor

\begin{equation}
    T \indices{^\mu _\nu} = \text{diag}(-\rho(t),p(t,r),p(t,r),p(t,r))
\end{equation}

\noindent where

\begin{equation}
    p(t,r) = \rho(t) \left( (w+1) \frac{2 r a(t) + M }{2 r a(t) - M}-1 \right)
\end{equation}

\noindent When $\nabla_\nu T^{\mu \nu}=0$ is solved, it gives the FLRW energy density $\rho(t)=\rho_0a(t)^{-3 (w+1)}$, where $w$ is the equation of state parameter for the FLRW background, and so we have $ \lim_{r \to \infty} p(t,r) \equiv p(t) = w \rho(t) $. The attractive feature here is that the energy density and, consequently, the $G_{00}$ field equations are the same as in the FLRW model. That is to say, the field equations asymptote to the FLRW field equations when $r \to \infty$ \cite{Gregoris_2020} which supports the idea that we are describing a central black hole embedded within a cosmological background. These quantities satisfy the GR field equations while leaving the scale factor $a(t)$ unconstrained. Typically, $a(t)$ is chosen from a corresponding FLRW solution. This allows us to study how a central inhomogeneity/black hole behaves within a background cosmology, or visa versa.

The black hole local horizon is, in general, dynamic. However, if $H(t) \to 0$ then the horizon corresponds to the Schwarzschild horizon, while if $H(t)=const$ then the horizon corresponds to the de-Sitter-Schwarzschild Horizon \cite{faraoni_2018}. 

\subsection{Perez and Romero}

In \cite{P_rez_2021}, Perez and Romero studied a generalized McVittie metric given by \eqref{mcvittie} with $k=0$ but with the function $m(t)$ undefined at the beginning of their analysis. This was for the purpose of finding a solution that has a black hole interacting with the cosmic fluid; i.e., this metric allows for accretion. The bounce mechanism comes from ad hoc quantum corrections to classical cosmology scale factor $a(t)$, essentially by solving the Wheeler-de Witt equation within the framework of de Broglie-Bohm quantum theory, which gives the following scale factor

\begin{equation}
    a(t) = a_b \bigg[ 1 + \left( \frac{t}{t_b} \right)^2 \bigg]^{\frac{1}{3}} \label{bounce}
\end{equation}

\noindent Notice, for $t \gg t_b$ (i.e., far away from the bounce), the scale factor reduces to that of FLRW with dust $a(t) \sim t^{\frac{2}{3}}$.

When we calculate the field equations, we get three independent field equations but we have five functions $a(t),m(t),q(t,r),\rho(t,r),$ and $P(t,r)$. Because of this, Perez and Romero choose to specify $a(t)$ and $m(t)$ \emph{a priori}, then solve for $q(t,r),\rho(t,r),$ and $P(t,r)$ trivially by defining these functions as the solutions since these three functions have no derivatives acting on them. In this sense their model is not an exact solution. In \cite{perez}, $m(t) = m_0$ was chosen. In \cite{P_rez_2021} they chose $m(t)=\frac{m_0}{a(t)}$, which is the choice made in the standard McVittie metric \cite{Mcvittieoriginal}. The former choice allows Perez and Romero to claim that the event horizon of the black hole is given by $R = 2 \frac{Gm_0}{c^2} a(t)$, where $R$ is the areal radius. With the latter choice, the black hole trapping horizon merges with the cosmological horizon, leading to a naked singularity.

\section{Black Hole Persistence: A Perturbative Scheme}

We seek to investigate the effect of black hole persistence in cosmology. The first requirement for such a study is to generate a bounce. Unlike some papers in the literature that use \emph{ad hoc} justifications and the fact that the McVittie solution is under-constrained to impose a bounce, we will generate it from the field equations of a modified theory of gravity. In  our current work we relax the assumptions of \cite{perez}. We obtain a positive curvature exact FLRW bouncing solution, where there are no assumptions on the form of the scale factor, derived from the field equations within the theory (so that no matter energy-conditions are violated). We then utilize the McVittie metric (actually its generalization) to model a central inhomogeneity embedded within a cosmology. As a consequence, the field equations will become increasingly complex which we solve by utilizing a perturbative scheme in which the central inhomogeneity is represented by a perturbation to the leading order bouncing FLRW solution.

To generate the bounce we will use New General Relativity (NGR) which is a teleparallel theory of gravity. This will allow us to utilize the simple bounce within GR for a closed ($k=+1$) FLRW model with a positive cosmological constant, while NGR will introduce a new parameter into the bounce solution to help mitigate the observational mismatch of the GR bounce. The methodology we use can be used for other modified theories of gravity. NGR was chosen as the setting to express this methodology due to its relative novelty in the literature and the complexity of its field equations which provides a good test-bed for the methodology.

The perturbative scheme that we use will be constructed in such a way that the ensuing solution is valid ``near the bounce". This means that the theory we choose will not be thought of as a full theory that applies globally in space or time, rather it will be thought of as an effective theory that is valid near the bounce in time (i.e., at very early times ) and far away from the central inhomogeneity in space due to the fact that the FLRW solution will dictate the leading order behaviour. Indeed, this analysis will be valid for \emph{any} theory that locally has an effective positive spatial curvature and an effective cosmological constant.

A number of alternative theories of gravity have been investigated due to cosmological tensions \cite{1} \cite{Valentino}. Observational data has been investigated in torsion theories  and bouncing models in torsion-based approaches such as teleparallel theories in \cite{Valentino} \cite{TorBounce}. In torsion theories of gravity the geometry gives rise to or augments a possible positive spatial curvature. NGR is a class of teleparallel theories of gravity where the principle geometric object is the torsion, which is determined from the co-frame and a flat, metric-compatible spin connection \cite{aldrovandi1732013}, which incorporates additional scalars constructed from the irreducible parts of the torsion in its action. To study spherically symmetric solutions in teleparallel geometry, it is necessary to first specify the general forms of the frame (or co-frame) and the flat, metric-compatible spin connection in the covariant approach \cite{krssak362019} that is compatible with the underlying spherically symmetric affine symmetries \cite{coley616439}.

Teleparallel geometry is defined by the tetrad $h \indices{^a _\mu}$ and a curvature free spin connection $\omega \indices{^a _b _\mu}$, where $a,b,...=1,2,3,4$ are the tangent space indices and $\mu,\nu,...=1,2,3,4$ are the space-time indices. We can then derive the metric by the relation

\begin{equation}
    g_{\mu \nu} = \eta_{ab} h \indices{^a _\mu} h \indices{^b _\nu} \label{telemetric}
\end{equation}

\noindent where $\eta_{ab} = \text{diag}(-1,1,1,1)$ is the tangent space Minkowski metric. We can then define the teleparallel connection by

\begin{equation}
    \Gamma \indices{^\rho _\nu _\mu} = h \indices{_a ^\rho} (\partial_\mu h \indices{^a _\nu} + \omega \indices{^a _b _\mu} h \indices{^b _\nu}) \label{tgconnection}
\end{equation}

\noindent The field strength of teleparallel gravity is defined by the torsion tensor

\begin{equation}
    T \indices{^\sigma _\mu _\nu} = 2 \Gamma \indices{^\sigma _{[\nu \mu]}}
\end{equation}

\noindent The three-parameter NGR is a teleparallel theory with the following action: 

\begin{equation}
    S = \frac{1}{2}\int d^4x h (c_a \mathcal{A} +c_t \mathcal{T}+ c_v \mathcal{V})
\end{equation}

\noindent where we are using $8 \pi G=c=1$ units, $h$ is the determinant of $h \indices{^a _\mu}$ and $\mathcal{A},\mathcal{T},\mathcal{V}$ are the scalars of the components of the torsion tensor that are irreducible under the Lorentz group defined by

\begin{eqnarray}
    \mathcal{A}_\mu &=& \frac{1}{6} \epsilon_{\mu \nu \rho \sigma} T^{\nu \rho \sigma} \\
    \mathcal{V}_\mu &=& T \indices{^\nu _\nu _\mu} \\
    \mathcal{T}_{\sigma \mu \nu} &=& T_{(\sigma \mu) \nu} + \frac{1}{3} (g_{\sigma [\nu} \mathcal{V}_{\mu]} + g_{\mu [\nu} \mathcal{V}_{\sigma]})
\end{eqnarray}

\noindent and

\begin{eqnarray}
    \mathcal{A} &=& \mathcal{A}^\mu \mathcal{A}_\mu \\
    \mathcal{V} &=& \mathcal{V}^\mu \mathcal{V}_\mu \\
    \mathcal{T} &=& \mathcal{T}^{\sigma \mu \nu} \mathcal{T}_{\sigma \mu \nu} 
\end{eqnarray}

\noindent Then we can write the torsion scalar $T$ as a combination of the above scalars

\begin{equation}
    T = \frac{3}{2}\mathcal{A}-\frac{2}{3}\mathcal{V}+\frac{2}{3}\mathcal{T}
\end{equation}

\noindent We use a normalization of this action that removes the cases when there is no TEGR limit, where the new constants have been defined as $b_1 = -\frac{2}{3}+c_t, b_2=3c_t,b_3=\frac{2}{3}-\frac{4}{9}c_a$. Notice, this implies that we now have two independent parameters $b_1$ and $b_3$ due to the relationship $b_2 = 2+ 3 b_1$. We can clearly see that the TEGR limit is when $b_1,b_3 \rightarrow 0$. Here, we consider the NGR theory described by $b_1=0$ which is known to be free of ghosts and have a weak-field Newtonian limit when $b_3 > 0$ \cite{diego}. Including a cosmological constant gives the full theory that we will consider

\begin{equation}
    S = \frac{1}{2}\int d^4x h (T - \tfrac{9}{4}b_3 \mathcal{A} - 2 \Lambda) + S^{(m)} \label{action}
\end{equation}

\noindent For a more detailed review of teleparallel gravity and NGR see \cite{Bahamonde_2023}.

In analogy to black hole apparent horizons, we will also use marginally trapped surfaces to define the location of the cosmological local horizon. We will begin by first deriving the bouncing NGR FLRW solution.

\subsection{NGR Bounce}

There exists a bouncing FLRW solution in GR that occurs with a radiation fluid, a positive cosmological constant, and positive spatial curvature. However, this bounce model in GR is not supported by current observations \cite{coleypersistance}. We consider a modified theory of gravity, namely NGR, with the goal of reproducing the same bounce but with new freedom coming from additional parameters introduced from NGR that may be compatible with observations of the early universe. Even so, this model will be considered an effective theory near the bounce rather than one that explains global cosmological behaviour. We can find a bouncing FLRW solution in NGR using the following procedure. Beginning with the general spherically symmetric tetrad

\begin{equation}
    h\indices{^a_\mu} = \text{diag}(A_1(t,r),A_2(t,r),A_3(t,r),A_3(t,r)\sin{\theta} ) \label{tetrad}
\end{equation}

\noindent we obtain the following spacetime metric using \eqref{telemetric}

\begin{equation}
    ds^2 = - A_1^2dt + A_2^2 dr^2 + A_3^2 ( d \theta^2 +  \sin^2\theta d \phi^2)
\end{equation}

\noindent The non-zero components of the spin connection $\omega_{abc}$ ($\omega_{abc} = - \omega_{bac}$) are given by 

\begin{align}
    \omega_{341} =W_1(t,r), \quad \omega_{342} = W_2(t,r), \quad \omega_{233} = \omega_{244} = W_3(t,r) \nonumber \\
    \omega_{234} = -\omega_{243} = W_4(t,r), \quad \omega_{121} = W_5(t,r), \quad \omega_{122} = W_6(t,r) \\
    \omega_{133} = \omega_{144} = W_7(t,r), \quad \omega_{134} = -\omega_{143} = W_8(t,r), \quad \omega_{344} = - \frac{\cot \theta }{A_3 \sin \theta} \nonumber
\end{align}

\noindent where

\begin{eqnarray}
    W_1 = -\frac{\dot{\chi}}{A_1}, \quad W_2 = -\frac{\chi'}{A_2}, \quad W_3 = \frac{\cosh \psi \cos \chi}{A_3}, \quad W_4 = \frac{\cosh \psi \sin \chi}{A_3} \\
    W_5 = -\frac{\dot{\psi}}{A_1}, \quad W_6 = -\frac{\psi'}{A_2}, \quad W_7 = \frac{\sinh \psi \cos \chi}{A_3}, \quad W_8 = \frac{\sinh \psi \sin \chi}{A_3}
\end{eqnarray}

\noindent We choose the FLRW metric in isotropic coordinates (to correspond to the McVittie metric which we will consider later)

\begin{equation}
    A_1 = 1, \quad A_2  = \frac{a(t)}{1+\frac{k}{4}r^2}, \quad A_3 = r A_2
\end{equation}

\noindent with a perfect fluid source

\begin{equation}
    \Theta \indices{^\mu _\nu} = \text{diag}(- \rho,p,p,p)
\end{equation}

\noindent We choose the following functions such that they satisfy the antisymmetric field equations of NGR FLRW for $k \geq 0$:

\begin{equation}
    \psi = 0, \quad \chi = \cos^{-1}\left(\frac{-4+kr^2}{4+kr^2} \right) \label{psichiFLRW}
\end{equation}

\noindent We can then calculate the NGR field equations with the normalized Lagrangian discussed above with $b_1=0$ and a cosmological constant $\Lambda > 0$. The nonzero components are

\begin{eqnarray}
    W \indices{^0 _0} &:& -\frac{(6-9 b_3)k}{2 a^2}- 3 \frac{\dot{a}^2}{a^2} + \Lambda = -\rho \label{fried1} \\
    W \indices{^i _j} &:& -\frac{(2-3 b_3)k}{2a^2} - \frac{\dot{a}^2}{a^2}-\frac{2 \ddot{a}}{a} + \Lambda = p \label{accel1}
\end{eqnarray}

\noindent We define $B \equiv (1-\frac{3}{2} b_3)k$. Then, as in GR, we obtain the two familiar equations 

\begin{eqnarray}
    \frac{\dot{a}^2}{a^2}  &=& \frac{\rho}{3} + \frac{\Lambda}{3} - \frac{B}{a^2} \label{fried2}  \\
    \frac{\ddot{a}}{a} &=& \frac{\Lambda}{3}-\frac{1}{6} (\rho +3p) \label{accel2}
\end{eqnarray}

\noindent Notice that these are the same equations as in GR, except with the modification $k \rightarrow B \equiv (1-\frac{3}{2} b_3)k$. This means we have the same solution for a closed universe cosmological bounce with radiation fluid, $\rho = \rho_0/a^4$, and a cosmological constant, but with more freedom in the constants due to the appearance of $b_3$. The solution is given by

\begin{equation}
    a(t) = \sqrt{\frac{a_-^2 \tanh^2 \left(\sqrt{\frac{\Lambda}{3}}(t+c) \right)-a_+^2}{\tanh^2 \left(\sqrt{\frac{\Lambda}{3}}(t+c) \right)-1}} \label{ngrbouncesoln}
\end{equation}

\noindent where

\begin{equation}
    a_\pm^2 = \frac{3}{2 \Lambda} \left( B \pm \sqrt{B^2 - \frac{4}{9}\Lambda \rho_0} \right) \label{apm}
\end{equation}

\noindent with the following conditions for a bounce to occur

\begin{equation}
    B \equiv ( 1-\tfrac{3}{2}b_3)k > 0, \quad B^2 \equiv  ( 1-\tfrac{3}{2}b_3)^2k^2 \geq \tfrac{4}{9}\Lambda \rho_r
\end{equation}

\noindent and $c$ is a constant which we will choose to be $c=0$ so that the minimum of the bounce occurs at $t=0$. 

Note that this is the bounce solution for $B>0$ and $\Lambda>0$ with radiation fluid. To obtain the classic big crunch closed universe solution under the same conditions, simply swap $a_-^2$ with $a_+^2$. To recover the GR bouncing solution we simply need to set $B=1$ or $b_3 = 0$. In GR, for the bounce to occur, we require $k=+1$, i.e., a closed universe. In NGR this requirement becomes $B>0$, which is satisfied by $k=+1$ with $b_3 < \frac{2}{3}$. We note that the perturbations cannot model a central inhomogeneity when $k=-1$ due to choice \eqref{psichiFLRW}. The minimum of the bounce occurs at $a(t) = a_+$. Indeed, we have the freedom to change the minimum of the scale factor by adjusting $b_3$, which can potentially allow this bounce mechanism to better fit the observations compared to the GR case.

\subsection{Setup of Perturbative Scheme}

Our goal is to find a bouncing cosmological solution with a central inhomogeneity in NGR. This is achieved by beginning with a FLRW solution, where the bounce is generated, at least in part, by the NGR contributions. However, due to the complexity of the NGR field equations, we will choose to employ perturbative methods to find such a solution. Using the idea that the McVittie metric models a central inhomogeneity within a cosmological background, combined with its FLRW limit when the mass function goes to zero, we will consider the following generalized McVittie metric

\begin{equation}
    ds^2 = - \left( \frac{ 1-\frac{M(t,r)}{2r}\omega(r) }{1+\frac{M(t,r)}{2r}\omega(r)} \right)^2 dt^2+ A(t,r)^2 \frac{\left( 1+\frac{M(t,r)}{2r}\omega(r) \right)^4}{\omega(r)^4} (dr^2 + r^2 d\Omega^2) \label{genmetric}
\end{equation}

\noindent where $\omega(r) \equiv \sqrt{1+\frac{k}{4}r^2}$, and the energy-momentum tensor is given by

\begin{equation}
    \Theta \indices{^\mu _\nu} = \text{diag} (-\rho(t,r), p_r(t,r), p_t(t,r), p_t(t,r))
\end{equation}

\noindent Notice, we have further generalized the McVittie metric by allowing the metric functions $A$ and $M$ to be both functions of $t$ and $r$. This is not the most general metric that can possibly describe a black hole within a cosmology, however, the analysis in this paper will be limited to the generalized McVittie metric. In general, the energy-momentum tensor models an inhomogeneous imperfect fluid with radial and tangential pressures, but we will pursue a perfect fluid solution later on. This allows us to potentially find solutions that are more general than even the generalizations of the original McVittie metric used in the literature with a general mass function $M(t)$, but still inherit some of the attractive features. For example, this metric is spherically symmetric and has the FLRW limit when $M(t,r) \to 0$ and $A(t,r) \to A(t)$. In fact, the solution we find will also have the feature that when $r \to 2$ the metric tends towards FLRW (i.e., is spatially homogeneous). Note, as mentioned before, that in the $k=+1$ case the limit $r \to \infty$ is not the ``far from the center" limit, however, $r \to 2$ is also not exactly a ``weak field limit" rather it is the farthest point from the center within a $k=+1$ closed universe. In the following text, we will use take $r \to 2$ to be far away from the central inhomogeneity and hence in the FLRW regime, which will be clarified in a later section. This naturally motivates the interpretation that this solution models a FLRW background with a central inhomogeneity.

To this end, we will choose the following perturbative ansatz

\begin{eqnarray}
    A(t,r) &=& a(t) + \epsilon \tilde{a}(t,r) \label{asymansatzA} \\
    M(t,r) &=& 0 + \epsilon \tilde{m}(t,r) \\
    \rho(t,r) &=& \frac{\rho_0}{a(t)^4} + \epsilon \tilde{\rho}(t,r) \\
    p_r(t,r) &=& \frac{\rho_0}{3a(t)^4} + \epsilon \tilde{p}_r(t,r) \\
    p_t(t,r) &=& \frac{\rho_0}{3a(t)^4} + \epsilon \tilde{p}_t(t,r) \\ 
    \psi(t,r) &=& 0 + \epsilon \tilde{\psi}(t,r) \\
    \chi(t,r) &=& \cos^{-1}\left(\frac{-4+k r^2}{4+k r^2} \right) + \epsilon \tilde{\chi}(t,r) \label{asymansatzchi}
\end{eqnarray}

\noindent Here $\epsilon$ is a dummy asymptotic parameter, it only keeps track of the order of the terms and will at the end be set to $\epsilon=1$. Notice that when $\epsilon=0$, the metric \eqref{genmetric} becomes FLRW in isotropic coordinates with a cosmological constant and radiation fluid. We will then expand the NGR field equations of \eqref{genmetric} in terms of $\epsilon$ up to $\epsilon^1$ order using the perturbed ansatz \eqref{asymansatzA}-\eqref{asymansatzchi}. For simplicity all terms of the field equations will be put on the left hand side. That is, 

\begin{equation}
    E_{\mu \nu} \equiv W_{\mu \nu} + \Lambda  g_{\mu \nu} - \Theta_{\mu \nu}
\end{equation}

\noindent then $E_{\mu \nu}=0$ expresses the field equations.
\begin{eqnarray}
    E_{\mu \nu}=E^{(0)}_{\mu \nu} + \epsilon E^{(1)}_{\mu \nu} + \mathcal{O}(\epsilon^2) = 0
\end{eqnarray}

\noindent Then we will solve $E^{(i)}_{\mu \nu}=0$ for each order up to $\epsilon^1$ ($i=0,1$) omitting all $E^{(i)}_{\mu \nu}$ for $i \geq 2$ as negligible. We have actually chosen the leading order ($\epsilon^0$) asymptotic ansatz \eqref{asymansatzA}-\eqref{asymansatzchi} such that $E^{(0)}_{\mu \nu}=0$ are satisfied with the NGR FLRW bounce solution discussed in the previous section,

\begin{equation}
    a(t) = \sqrt{\frac{a_-^2 \tanh^2 \left(\sqrt{\frac{\Lambda}{3}}t\right)-a_+^2 }{ \tanh^2 \left(\sqrt{\frac{\Lambda}{3}}t\right)-1 }} \label{flrwsoln}
\end{equation}

\noindent As $t \to 0$, which is when the bounce occurs,

\begin{equation}
    a(t) = a_+ + \frac{(a_+^2 - a_-^2) \Lambda}{6 a_+}t^2 + \frac{(a_+^2 - a_-^2)(a_+^2 + 3a_-^2) \Lambda^2}{216 a_+^3} t^4 + \mathcal{O}(t^6) \label{ascsolnpert}
\end{equation}

\noindent All that is left is to solve for the corrections to the NGR FLRW solution by solving $E^{(1)}_{\mu \nu}=0$ for the functions in the asymptotic ansatz with a tilde.

\subsection{Field Equations}

First, we list the non-zero components of $E^{(1)}_{\mu \nu}=0$.

\begin{multline}
    E^{(1)}_{(11)} = \frac{1}{16 a^3 r} \bigg[ 16 a^3 r \tilde{\rho}-48 \dot{a} a^2 \dot{\tilde{m}} \sqrt{k r^2+4}+a (r (-96 \dot{a} \dot{\tilde{a}}-12 b_3 \sqrt{k} (k r^2+4) \tilde{\chi}' \\ +k r (k r^2+8) \sqrt{k r^2+4} \tilde{m}'') -12 \tilde{m} \sqrt{k r^2+4} (4 \dot{a}^2+(6 b_3-5) k)+24 b_3 \sqrt{k} \tilde{\chi} (k r^2-4)\\ +16 \sqrt{k r^2+4} \tilde{m}'')+48 r \tilde{a} (2 \dot{a}^2+(2-3 b_3) k)  +2 (k r^2+4) (r (k r^2+4) \tilde{a}''+8 \tilde{a}') \bigg] \label{E(11)}
\end{multline}

\begin{multline}
    E^{(1)}_{(22)} = \frac{1}{8 a^3 r} \bigg[-8r\tilde{a}((3 b_3-2) k-2 \dot{a}^2-2a\ddot{a})+(16-k^2 r^4) \tilde{a}'+2 a (r (b_3 \sqrt{k} (k r^2+4) \tilde{\chi}'-8 \dot{a} \dot{\tilde{a}}) \\ -2 \tilde{m} \sqrt{k r^2+4} (2 \dot{a}^2+(3 b_3-2) k)+4 b_3 \sqrt{k} \tilde{\chi} (k r^2-4))-16a^2(\sqrt{k r^2+4}\tilde{m}\ddot{a}+2 \dot{a} \dot{\tilde{m}} \sqrt{k r^2+4}+r\ddot{\tilde{a}}) \\ -8a^3(r \tilde{p}_r+\sqrt{k r^2+4}\ddot{\tilde{m}}) \bigg] \label{E(22)}
\end{multline}

\begin{multline}
    E^{(1)}_{(33)} = \frac{1}{16 a^3 r \sqrt{k r^2+4}} \bigg[16r\sqrt{k r^2+4}\tilde{a}((2-3 b_3) k+2 \dot{a}^2+2a\ddot{a})+(k r^2+4)^{5/2} (r \tilde{a}''+\tilde{a}') \\ -4 a \bigg(2 r \big(4 \dot{a} \dot{\tilde{a}} \sqrt{k r^2+4}+b_3 (k^{\frac{3}{2}} r^2 \sqrt{k r^2+4}+4 \sqrt{k (k r^2+4)} ) \tilde{\chi}' \big)+2 \tilde{m} (k r^2+4) (2 \dot{a}^2+(3 b_3-2) k) \\ +b_3 \tilde{\chi} (4 \sqrt{k (k r^2+4)}-k^{\frac{3}{2}} r^2 \sqrt{k r^2+4}) \bigg)-32a^2(k r^2+4\tilde{m}\ddot{a}+2 \dot{a} \dot{\tilde{m}} (k r^2+4)+r\sqrt{k r^2+4}\ddot{\tilde{a}}) \\ -16a^3(r \tilde{p}_t \sqrt{k r^2+4}+k r^2+4\ddot{\tilde{m}}) \bigg] \label{E(33)}
\end{multline}

\begin{multline}
    E^{(1)}_{(12)} = \frac{1}{a^2 r^2 (k r^2+4)^{\frac{3}{2}}} \bigg[a^2 \bigg((k r^2+4) (r (k r^2+4) \dot{\tilde{m}}'-4 \dot{\tilde{m}})-8 b_3 r^2 \dot{\tilde{\chi}} \sqrt{k (k r^2+4)} \bigg) \\ +a \bigg( (k r^2+4) (r \dot{a} (k r^2+4) \tilde{m}'+2 r^2 \sqrt{k r^2+4} \dot{\tilde{a}}'-4 \dot{a} \tilde{m})+4 b_3 k r^2 \tilde{\psi} \sqrt{k r^2+4} \bigg)-2 \dot{a} r^2 (k r^2+4)^{\frac{3}{2}} \tilde{a}' \bigg] \label{E(12)}
\end{multline}

\begin{equation}
    E^{(1)}_{[12]} = -\frac{4 b_3 (a \sqrt{k} \dot{\tilde{\chi}}+k \tilde{\psi})}{a (k r^2+4)} \label{E[12]}
\end{equation}

\begin{multline}
    E^{(1)}_{[34]} = \frac{b_3\sin (\theta)}{32 a^3 r^2} \bigg[ -12 \sqrt{k} r^2 (k r^2+4) \tilde{a}'+a \bigg(r (64 \dot{a} \sqrt{k} r \tilde{\psi}+(k r^2+4) (r (k r^2+4) \tilde{\chi}''+8 \tilde{\chi}')) \\ -2 \tilde{\chi} (k^2 r^4-40 k r^2+16) \bigg)+16 a^2 r^2 (2 \sqrt{k} \dot{\tilde{\psi}}-3 \dot{a} \dot{\tilde{\chi}})-16a^3r^2\ddot{\tilde{\chi}} \bigg] \label{E[34]}
\end{multline}

\noindent Here, a prime indicates a partial derivative with respect to the radial coordinate and a dot with respect to the time coordinate. First, we solve \eqref{E[12]} for $\tilde{\psi}$ by

\begin{equation}
    \tilde{\psi} = -\frac{a \dot{\tilde{\chi}}}{\sqrt{k}} \label{solvetpsi}
\end{equation}

\noindent Then we notice that \eqref{E(11)} - \eqref{E(33)} can be algebraically solved for $\tilde{\rho}, \tilde{p}_r$ and $\tilde{p}_t$, respectively, since derivatives of these functions do not appear in the corresponding equations. We are left with three unknown functions $\tilde{a}, \tilde{m}$ and $\tilde{\chi}$, but only two equations \eqref{E(12)} and \eqref{E[34]}. To fix this, we assert the additional constraint that the $\epsilon^1$ order corrections to the tangential and radial pressures are equal, $\tilde{p}_t-\tilde{p}_r = 0$, i.e., the matter is a perfect fluid at the $\epsilon^1$ order, which yields

\begin{equation}
    \frac{12 a b_3 \sqrt{k} \bigg(r \left(k r^2+4\right) \tilde{\chi}'+\tilde{\chi} \left(k r^2-4\right)\bigg)+\left(k r^2+4\right) \bigg(\left(4-3 k r^2\right) \tilde{a}'-r \left(k r^2+4\right) \tilde{a}''\bigg)}{16 a^3 r} =0 \label{presconstraint}
\end{equation}

\noindent After substituting \eqref{solvetpsi} into \eqref{E(12)} and \eqref{E[34]} we have, respectively,

\begin{multline}
    \frac{\left(k r^2+4\right) \left(r \left(k r^2+4\right) \dot{\tilde{m}}'-4 \dot{\tilde{m}}\right)-12 b_3 r^2 \dot{\tilde{\chi}} \sqrt{k \left(k r^2+4\right)}}{r^2 \left(k r^2+4\right)^{\frac{3}{2}}} \\ -\frac{2 \dot{a} \tilde{a}'}{a^2}+\frac{r \left(\dot{a} \left(k r^2+4\right) \tilde{m}'+2 r \sqrt{k r^2+4} \dot{\tilde{a}}'\right)-4 \dot{a} \tilde{m}}{a r^2 \sqrt{k r^2+4}} = 0 \label{E(12)v2}
\end{multline}

\begin{multline}
    -\frac{b_3\sin(\theta)}{32 a^3 r^2} \bigg(12 \sqrt{k} r^2 (k r^2+4) \tilde{a}'+a \big(2 \tilde{\chi} (k^2 r^4-40 k r^2+16)-r (k r^2+4) (r (k r^2+4) \tilde{\chi}''+8 \tilde{\chi}') \big) \\ +144 a^2 \dot{a} r^2 \dot{\tilde{\chi}}+48a^3r^2\ddot{\tilde{\chi}} \bigg) = 0 \label{E[34]v2}
\end{multline}

\noindent We are then left to solve  for $\tilde{a}, \tilde{m}$ and $\tilde{\chi}$ using equations \eqref{presconstraint}, \eqref{E(12)v2}, and \eqref{E[34]v2}.

\subsection{Second Perturbation}

We care about a solution only near the bounce (that is to say near $t=0$). Hence, we will attempt  to solve these differential equations by writing $\tilde{a}, \tilde{m}$ and $\tilde{\chi}$ as a series in time as follows:

\begin{eqnarray}
    \tilde{a}(t,r) &=& \tilde{a}_0(r)+\tilde{a}_1(r) t +\tilde{a}_2(r) t^2 + \tilde{a}_3(r) t^3 \\
    \tilde{m}(t,r) &=& \tilde{m}_0(r)+\tilde{m}_1(r) t +\tilde{m}_2(r) t^2 + \tilde{m}_3(r) t^3 \\
    \tilde{\chi}(t,r) &=& \tilde{\chi}_0(r)+\tilde{\chi}_1(r) t +\tilde{\chi}_2(r) t^2 + \tilde{\chi}_3(r) t^3
\end{eqnarray}

\noindent Note that the three equations only contain first derivatives in time except for \eqref{E[34]v2} which contains second time derivatives of $\tilde{\chi}$; hence, to be able to solve these equations as a series solution that is valid up to $t^2$ order while maintaining full generality, we would need to expand $\tilde{\chi}$ up to $t^4$ order and only solve the equations up to and including the $t^2$ order. However, as we shall see later, the complexity of the equations prevent us from doing so. This does not disqualify the following solution, rather this solution is not necessarily the most general one. However, this caveat will not be significant, as the results of this section will motivate an exact ansatz as will be discussed in the next section.

In addition, we will need to replace the leading order scale factor function $a(t)$ appearing in the equations with the series expansion of \eqref{flrwsoln} up to order $t^4$ given by \eqref{ascsolnpert}. After making these substitutions into \eqref{presconstraint}, \eqref{E(12)v2} and \eqref{E[34]v2}, we then expand in terms of $t$ up to order $t^3$ to obtain the following equations:

\begin{multline}
    \bigg[  a_+^2 (k r^2+4) (a_+ r (k r^2+4) \tilde{m}_1'(r)-4 a_+ \tilde{m}_1(r)+2 r^2 \sqrt{k r^2+4} \tilde{a}_1'(r))  \\  -12 a_+ b_3 r^2 \sqrt{k (k r^2+4)} \tilde{\chi}_1(r)\bigg] +  \\  t \bigg[\frac{1}{3} (k r^2+4) (r (a_+ (12 a_+ r \sqrt{k r^2+4} \tilde{a}_2'(r)-(k r^2+4) (\Lambda  (a_-^2-a_+^2) \tilde{m}_0'(r)-6 a_+^2 \tilde{m}_2'(r)))  \\  +2 \Lambda  r (a_-^2-a_+^2) \sqrt{k r^2+4} \tilde{a}_0'(r))+4 a_+ \Lambda  (a_-^2-a_+^2) \tilde{m}_0(r)-24 a_+^3 \tilde{m}_2(r))  \\  -24 a_+^3 b_3 r^2 \sqrt{k (k r^2+4)} \tilde{\chi}_2(r) \bigg] +  \\  t^2 \bigg[\frac{1}{3} (k r^2+4) (r (a_+ (18 a_+ r \sqrt{k r^2+4} \tilde{a}_3'(r)-(k r^2+4) (2 \Lambda  (a_-^2-a_+^2) \tilde{m}_1'(r)-9 a_+^2 \tilde{m}_3'(r)))  \\  +\Lambda  r (a_-^2-a_+^2) \sqrt{k r^2+4} \tilde{a}_1'(r))+8 a_+ \Lambda  (a_-^2-a_+^2) \tilde{m}_1(r)-36 a_+^3 \tilde{m}_3(r)) \\ +4 a_+ b_3 \Lambda  r^2 (a_-^2-a_+^2) \sqrt{k (k r^2+4)} \tilde{\chi}_1(r)-36 a_+^3 b_3 r^2 \sqrt{k (k r^2+4)} \tilde{\chi}_3(r) \bigg] \\ \frac{t^3}{27 a_+^2} \bigg[ \Lambda   (a_-^2-a_+^2) ((k r^2+4) (r (\Lambda  r (3 a_-^2+a_+^2) \sqrt{k r^2+4} \tilde{a}_0'(r)-a_+^3 (k r^2+4) (2 \Lambda  \tilde{m}_0'(r)+27 \tilde{m}_2'(r))) \\ +8 a_+^3 \Lambda  \tilde{m}_0(r)+108 a_+^3 \tilde{m}_2(r))+216 a_+^3 b_3 r^2 \sqrt{k (k r^2+4)} \tilde{\chi}_2(r)) \bigg] + \mathcal{O}(t^4) = 0
\end{multline}

\begin{multline}
   \bigg[ a_+ (r ((k r^2+4) (a_+ r (k r^2+4) \tilde{\chi}_0''(r)+8 a_+ \tilde{\chi}_0'(r)-12 \sqrt{k} r \tilde{a}_0'(r))-96 a_+^3 r \tilde{\chi}_2(r)) \\ -2 a_+ (k r^2 (k r^2-40)+16) \tilde{\chi}_0(r)) \bigg] + \\ t \bigg[ a_+ (r ((k r^2+4) (a_+ r (k r^2+4) \tilde{\chi}_1''(r)+8 a_+ \tilde{\chi}_1'(r)-12 \sqrt{k} r \tilde{a}_1'(r))-288 a_+^3 r \tilde{\chi}_3(r)) \\ -2 a_+ \tilde{\chi}_1(r) (-8 r^2 (3 \Lambda  (a_-^2-a_+^2) +5 k)+k^2 r^4+16)) \bigg] + \\ t^2 \bigg[ \frac{1}{6} (r (k r^2+4) (8 \Lambda  (a_+^2-a_-^2) \tilde{\chi}_0'(r)-a_-^2 k \Lambda  r^3 \tilde{\chi}_0''(r)-4 a_-^2 \Lambda  r \tilde{\chi}_0''(r)+a_+^2 k \Lambda  r^3 \tilde{\chi}_0''(r) \\ +6 a_+^2 r (k r^2+4) \tilde{\chi}_2''(r)+4 a_+^2 \Lambda  r \tilde{\chi}_0''(r)+48 a_+^2 \tilde{\chi}_2'(r)-72 a_+ \sqrt{k} r \tilde{a}_2'(r)) \\ -12 a_+^2 \tilde{\chi}_2(r) (-8 r^2 (9 \Lambda  (a_-^2-a_+^2) +5 k)+k^2 r^4+16) \\ +2 \Lambda  (a_-^2-a_+^2)  (k r^2 (k r^2-40)+16) \tilde{\chi}_0(r)) \bigg] + \\ t^3 \bigg[ \frac{1}{6} (-2 \Lambda  (a_-^2-a_+^2)  \tilde{\chi}_1(r) (8 r^2 (3 a_-^2 \Lambda -7 a_+^2 \Lambda +5 k)-k^2 r^4-16) \\ +r (k r^2+4) (8 \Lambda  (a_+^2-a_-^2) \tilde{\chi}_1'(r)-a_-^2 k \Lambda  r^3 \tilde{\chi}_1''(r)-4 a_-^2 \Lambda  r \tilde{\chi}_1''(r)+a_+^2 k \Lambda  r^3 \tilde{\chi}_1''(r)+6 a_+^2 r (k r^2+4) \tilde{\chi}_3''(r) \\ +4 a_+^2 \Lambda  r \tilde{\chi}_1''(r)+48 a_+^2 \tilde{\chi}_3'(r)-72 a_+ \sqrt{k} r \tilde{a}_3'(r))-12 a_+^2 \tilde{\chi}_3(r) (-8 r^2 (18 \Lambda  (a_-^2-a_+^2) +5 k) \\ +k^2 r^4+16)) \bigg] + \mathcal{O}(t^4) = 0
\end{multline}

\begin{multline}
    \bigg[-a_+ ((k r^2+4) (-12 a_+ b_3 \sqrt{k} r \tilde{\chi}_0'(r)+r (k r^2+4) \tilde{a}_0''(r)+(3 k r^2-4) \tilde{a}_0'(r)) \\ -12 a_+ b_3 \sqrt{k} (k r^2-4) \tilde{\chi}_0(r)) \bigg] \\  + t \bigg[-a_+ ((k r^2+4) (-12 a_+ b_3 \sqrt{k} r \tilde{\chi}_1'(r)+r (k r^2+4) \tilde{a}_1''(r)+(3 k r^2-4) \tilde{a}_1'(r)) \\ -12 a_+ b_3 \sqrt{k} (k r^2-4) \tilde{\chi}_1(r))\bigg] + \\ t^2 \bigg[ (-(k r^2+4) (2 b_3 \sqrt{k} r (\Lambda  (a_-^2-a_+^2)  \tilde{\chi}_0'(r)-6 a_+^2 \tilde{\chi}_2'(r))+a_+ r (k r^2+4) \tilde{a}_2''(r) \\ +a_+ (3 k r^2-4) \tilde{a}_2'(r))-2 b_3 \Lambda  \sqrt{k} (a_-^2-a_+^2)  (k r^2-4) \tilde{\chi}_0(r)+12 a_+^2 b_3 \sqrt{k} (k r^2-4) \tilde{\chi}_2(r)) \bigg] + \\ t^3 \bigg[ (-(k r^2+4) (2 b_3 \sqrt{k} r (\Lambda  (a_-^2-a_+^2)  \tilde{\chi}_1'(r)-6 a_+^2 \tilde{\chi}_3'(r))+a_+ r (k r^2+4) \tilde{a}_3''(r)+a_+ (3 k r^2-4) \tilde{a}_3'(r)) \\ -2 b_3 \Lambda  \sqrt{k} (a_-^2-a_+^2)  (k r^2-4) \tilde{\chi}_1(r)+12 a_+^2 b_3 \sqrt{k} (k r^2-4) \tilde{\chi}_3(r)) \bigg] + \mathcal{O}(t^4) = 0
\end{multline}

\noindent We then solve for the functions of $r$ at each of the powers of $t$, which gives the following solution:

\begin{eqnarray}
    \tilde{a}_0(r) &=& \frac{4 a_+^3 c_6-2 a_+ c_5 k}{k^{\frac{3}{2}} \left(k r^2+4\right)}+c_9 \label{perttildes1}\\
    \tilde{a}_1(r) &=& \frac{2 a_+^3 (c_1 \Lambda +6 c_2)-2 a_+ c_1 \left(a_-^2 \Lambda +k\right)}{k^{\frac{3}{2}} \left(k r^2+4\right)}+c_{10} \\
    \tilde{a}_2(r) &=& \frac{c_5 k \Lambda  (a_-^2-a_+^2)-6 a_+^2 c_6 (3 \Lambda  (a_-^2-a_+^2)+k)}{3 a_+ k^{\frac{3}{2}} \left(k r^2+4\right)}+c_{11}\\
    \tilde{a}_3(r) &=& \frac{c_1 \Lambda  (a_-^2-a_+^2) \left(3 a_-^2 \Lambda -7 a_+^2 \Lambda +3 k\right)-18 a_+^2 c_2 (6 \Lambda  (a_-^2-a_+^2)+k)}{9 a_+ k^{\frac{3}{2}} \left(k r^2+4\right)}+c_{12}\\
    \tilde{m}_0(r) &=& \frac{r \left(\frac{4 a_+^2 c_6 \left(2 \Lambda  \left(73 a_+^2-69 a_-^2\right)+27 (b_3-2) k\right)-4 c_5 k \Lambda  \left(3 a_-^2+a_+^2\right)}{k^{\frac{3}{2}} \Lambda  \left(9 a_-^2-5 a_+^2\right) \left(k r^2+4\right)}+c_7\right)}{\sqrt{k r^2+4}}\\
    \tilde{m}_1(r) &=& \frac{r \left(4 a_-^2 c_1 \Lambda -4 a_+^2 (c_1 \Lambda +6 c_2)+2 (2-3 b_3) c_1 k+c_3 k^{5/2} r^2+4 c_3 k^{\frac{3}{2}}\right)}{k^{\frac{3}{2}} \left(k r^2+4\right)^{\frac{3}{2}}}\\
    \tilde{m}_2(r) &=& \frac{r \left(\frac{2 c_5 k \Lambda  (a_-^2-a_+^2) \left(3 a_-^2+a_+^2\right)+12 a_+^2 c_6 \left(-a_-^4 \Lambda-a_+^2 \left(10 a_-^2 \Lambda +(3 b_3+4) k\right)+9 a_-^2 b_3 k+11 a_+^4 \Lambda \right)}{a_+^2 k^{\frac{3}{2}} \left(5 a_+^2-9 a_-^2\right) \left(k r^2+4\right)}+3 c_8\right)}{3 \sqrt{k r^2+4}} 
\end{eqnarray}

\begin{multline}
    \tilde{m}_3(r) = \frac{r}{9 a_+^2 k^{\frac{3}{2}} \left(k r^2+4\right)^{\frac{3}{2}}} \bigg[ 4 a_-^4 c_1 \Lambda ^2+2 a_-^2 \Lambda  \left(78 a_+^2 c_2+(2-3 b_3) c_1 k\right) \\ +a_+^2 \left(-4 a_+^2 \Lambda  (c_1 \Lambda +39 c_2)+2 (3 b_3-2) k (c_1 \Lambda -9 c_2)+9 c_4 k^{5/2} r^2+36 c_4 k^{\frac{3}{2}}\right) \bigg] \label{perttildes8}
\end{multline}
    
\begin{eqnarray}
    \tilde{\chi}_0(r) &=& \frac{c_5 r}{k r^2+4}, \quad \tilde{\chi}_1(r) = \frac{c_1 r}{k r^2+4} \\
    \tilde{\chi}_2(r) &=& \frac{c_6 r}{k r^2+4}, \quad \tilde{\chi}_3(r) = \frac{c_2 r}{k r^2+4} 
\end{eqnarray}

\noindent Note that for this simplified solution to be found we assumed the form of the $\tilde{\chi}_i$'s, then the rest of the functions were solved in general. We can see there are twelve constants $c_i$ that arise from integration. To reduce their number we need to assert some boundary conditions.

\section{Boundary Conditions}

We can apply boundary conditions motivated by the form of the McVittie metric and the goal to describe a central inhomogeneity. If we are modelling an FLRW background with a central inhomogeneity, then it is reasonable to assert the boundary condition that as we move farther from the central inhomogeneity the metric tends towards FLRW. However, we cannot simply take the limit $r \to \infty$ as the isotropic radial coordinate in a closed universe ($k=+1$) double covers the spacetime and the space itself is not infinite in distance. Meanwhile, in the open universe case ($k=-1$) we can get infinitely far away in terms of the areal radial coordinate, but in isotropic coordinates this infinity corresponds to $r = 2$. We can see this by transforming from the isotropic $r$ to the areal radial coordinate $R$:

\begin{equation}
    R = a(t) \bar{r} = \frac{a(t)r}{1+k\frac{r^2}{4}} \label{arealrad}
\end{equation}

\noindent where $\bar{r}$ is the co-moving radial coordinate. When $k=+1$ the range of values of the isotropic coordinate $r$, $\{ 0 < r < 2 \}$ and $ \{ 2 < r < \infty \}$ correspond to the same set of $R$ values, namely $\{ 0 < R < a(t) \}$ where the ``maximal" $R$ value is obtained in the limit $r \to 2$. And in the case of $k=-1$, $R$ tends to infinity when $r \to 2$, which means the only allowed values of $r$ are $0 < r < 2$. Regardless of the choice of $k$, the ``farthest" in terms of the areal radius $R$ we can be from the central inhomogeneity is at $r = 2$. Due to the difference between these coordinates for different choice of $k$, we will treat the boundary conditions separately.

\subsection{k = +1}

Since $R$ is always finite, we cannot assert the FLRW limit with $R \to \infty$. However, we can still assert the spirit of said limit. What the FLRW limit represents is the idea that the influence of a finite object diminishes as we move farther from the object until the influence vanishes when infinitely far away. Since we cannot reach infinitely far away, we will assert the condition that the influence of the central inhomogeneity must be \emph{decreasing} at the farthest point, $r = 2$. Note that we formulated an ansatz where the leading order behaviour is FLRW, this implicitly made the assumption that we are far away from the central inhomogeneity. Therefore, in the case of $k=+1$, the solution we found in the previous section is only valid in a neighbourhood of $r=2$. This means that the decreasing behaviour that we want to assert need only apply near $r=2$ rather than for all allowed values of $r$. For example, if we want to check if the function $\frac{r}{(4+r^2)^{3/2}}$ obeys this condition, we only need to check that this function is decreasing in a neighbourhood of $r=2$ and not be concerned with the fact that it is increasing for $0 < r < \sqrt{2}$ as this is outside of the region where the solution is valid.

Looking at the solution \eqref{perttildes1}-\eqref{perttildes8}, we see that when $k=+1$ all the functions of $r$ are of the following form

\begin{equation}
    f(r) = \frac{r^a}{(4+r^2)^b}
\end{equation}

\noindent multiplied by some constant factor, for $a,b \geq 0$. When $a < b$, $f(r)$ is decreasing near $r = 2$ and that this condition alone is enough to conclude that there exists some neighbourhood of $r=2$ such that if $a<b$, then in that neighbourhood $f(r)$ is decreasing: $f'(r)<0$. Therefore, the conditions $a<b$ satisfies the ``decreasing far away from the center" condition for all functions $f(r)$ of this form which is the case for all the terms in the solution. This encapsulates the ``FLRW limit" in a finite universe.

As mentioned before, the FLRW limit of the generalized McVittie metric \eqref{genmetric} is $A(t,r) \to A(t)$ and $M(t,r) \to 0$, which in our ansatz \eqref{asymansatzA}-\eqref{asymansatzchi} is equivalent to $\tilde{a} \to 0$ and $\tilde{m} \to 0$.  However, there is no reason that $\tilde{\chi}$ needs to obey this condition as it does not contribute to the metric. Therefore, we will refer to the ``FLRW limit" in the $k=+1$ case as the condition that the radial parts of the perturbations that represent the effects from the central inhomogeneity are decreasing near $r=2$, or simply, all terms in $\tilde{a}_i(r)$'s and $\tilde{m}_i(r)$'s \eqref{perttildes1}-\eqref{perttildes8} must obey $a < b$. For example, if we want to assert this FLRW limit on the function
\begin{equation}
    \frac{\lambda_0}{4+r^2} + \frac{\lambda_1 r^2}{4+r^2}
\end{equation}

\noindent then the first term obeys the condition, while the second does not. Therefore, requiring this condition be obeyed implies $\lambda_1=0$.

Applying this boundary condition to the solution \eqref{perttildes1}-\eqref{perttildes8} eliminates eight of the twelve constants of integration by requiring the following

\begin{equation}
    c_i = 0 \quad \forall i \neq 1,2,5,6.
\end{equation}

\noindent We then obtain

\begin{eqnarray}
    \tilde{a}_0(r) &=& \frac{4 a_+^3 c_6-2 a_+ c_5 k}{k^{\frac{3}{2}} \left(k r^2+4\right)}\\
    \tilde{a}_1(r) &=& \frac{2 a_+^3 (c_1 \Lambda +6 c_2)-2 a_+ c_1 \left(a_-^2 \Lambda +k\right)}{k^{\frac{3}{2}} \left(k r^2+4\right)}\\
    \tilde{a}_2(r) &=& \frac{c_5 k \Lambda  (a_-^2-a_+^2)-6 a_+^2 c_6 (3 \Lambda  (a_-^2-a_+^2)+k)}{3 a_+ k^{\frac{3}{2}} \left(k r^2+4\right)}\\
    \tilde{a}_3(r) &=& \frac{c_1 \Lambda  (a_-^2-a_+^2) \left(3 a_-^2 \Lambda -7 a_+^2 \Lambda +3 k\right)-18 a_+^2 c_2 (6 \Lambda  (a_-^2-a_+^2)+k)}{9 a_+ k^{\frac{3}{2}} \left(k r^2+4\right)}\\
    \tilde{m}_0(r) &=& r \left(\frac{4 a_+^2 c_6 \left(2 \Lambda  \left(73 a_+^2-69 a_-^2\right)+27 (b_3-2) k\right)-4 c_5 k \Lambda  \left(3 a_-^2+a_+^2\right)}{k^{\frac{3}{2}} \Lambda  \left(9 a_-^2-5 a_+^2\right) \left(k r^2+4\right)^{\frac{3}{2}}}\right)\\
    \tilde{m}_1(r) &=& \frac{r \left(4 a_-^2 c_1 \Lambda -4 a_+^2 (c_1 \Lambda +6 c_2)+2 (2-3 b_3) c_1 k\right)}{k^{\frac{3}{2}} \left(k r^2+4\right)^{\frac{3}{2}}}
\end{eqnarray}

\begin{multline}
    \tilde{m}_2(r) = \frac{r}{3a_+^2 k^{\frac{3}{2}} \left(5 a_+^2-9 a_-^2\right) \left(k r^2+4\right)^{\frac{3}{2}}} \bigg[ 2 c_5 k \Lambda  (a_-^2-a_+^2) \left(3 a_-^2+a_+^2\right) \\ +12 a_+^2 c_6 \left(-a_-^4 \Lambda-a_+^2 \left(10 a_-^2 \Lambda +(3 b_3+4) k\right)+9 a_-^2 b_3 k+11 a_+^4 \Lambda \right) \bigg]
\end{multline}

\begin{multline}
  \tilde{m}_3(r) = \frac{r}{9 a_+^2 k^{\frac{3}{2}} \left(k r^2+4\right)^{\frac{3}{2}}} \bigg[ 4 a_-^4 c_1 \Lambda ^2+2 a_-^2 \Lambda  \left(78 a_+^2 c_2+(2-3 b_3) c_1 k\right) \\ +a_+^2 \left(-4 a_+^2 \Lambda  (c_1 \Lambda +39 c_2)+2 (3 b_3-2) k (c_1 \Lambda -9 c_2)\right) \bigg]
\end{multline}
    
\begin{eqnarray}
    \tilde{\chi}_0(r) &=& \frac{c_5 r}{k r^2+4}, \quad \tilde{\chi}_1(r) = \frac{c_1 r}{k r^2+4}\\
    \tilde{\chi}_2(r) &=& \frac{c_6 r}{k r^2+4}, \quad \tilde{\chi}_3(r) = \frac{c_2 r}{k r^2+4}
\end{eqnarray}

\noindent We can immediately see that this series solution suggests that the functions $\tilde{a}, \tilde{m}$, and $\tilde{\chi}$ have the separable form

\begin{eqnarray}
    \tilde{a}(t,r) &=& \frac{1}{4+ kr^2} \alpha(t) \label{simpleansatz1} \\
    \tilde{m}(t,r) &=& \frac{r}{(4+kr^2)^{\frac{3}{2}}} \beta(t) \label{simpleansatz2}\\
    \tilde{\chi}(t,r) &=& \frac{r}{4 + kr^2} \gamma(t) \label{simpleansatz3}
\end{eqnarray}

\noindent for some functions of time $\alpha, \beta, \gamma$. In principle, this allows for the possibility of finding a solution without using a series in time; though we shall not take this approach here, it will still dramatically simplify the form of the analysis. 

\subsection{k = -1}

Note that the limit $R \to \infty$ does exist in the $k=-1$ case and is expressed in isotropic coordinates by $r \to 2^-$. However, assuming \eqref{psichiFLRW} (which is not strictly valid in the $k=-1$ case \cite{McNutt:2023nxm}) the functions are now of form $g(r) = \frac{r^a}{(4-r^2)^b}$ and are all increasing in the limit $r \to 2^-$. Therefore, all terms of the form $g(r)$ must be set to zero to obey the ``FLRW limit" unless $b = 0$. This yields

\begin{eqnarray}
    \tilde{a}_0(r) = c_9, \quad \tilde{a}_1(r) = c_{10}, \quad \tilde{a}_2(r) = c_{11}, \quad \tilde{a}_3(r) = c_{12}, \quad \tilde{m}_0(r) = \tilde{m}_1(r) = \tilde{m}_2(r) = \tilde{m}_3(r) = 0
\end{eqnarray}

\noindent This cannot represent FLRW with a central inhomogeneity because $\tilde{m}=0$ means the central inhomogeneity has vanished. Therefore, we conclude that the bouncing NGR FLRW solution \eqref{perttildes1}-\eqref{perttildes8} can only contain a central inhomogeneity in the case $k=+1$.

\subsection{New Ansatz}

We will now substitute \eqref{simpleansatz1}-\eqref{simpleansatz3} into \eqref{presconstraint}, \eqref{E(12)v2}, and \eqref{E[34]v2}. Immediately, this ansatz satisfies the requirement that the two ``tilde pressures" be equal, $\tilde{p}_r = \tilde{p}_t$ \eqref{presconstraint}, so from here on we will simply refer to this pressure as $\tilde{p}$. Then \eqref{E(12)v2} and \eqref{E[34]v2} become

\begin{eqnarray}
    -\frac{2r \left(6 a^2 b_3 \sqrt{k}  \dot{\gamma}+a k  \left(a \dot{\beta}+\dot{a} \beta+2 \dot{\alpha}\right)-2 \dot{a} k \alpha\right)}{a^2 \left(k r^2+4\right)^2} = 0 \label{E(12)simple} \\
    \frac{3 b_3 r \sin (\theta ) \left(-2 a^2 \left(a \ddot{\gamma}+3 \dot{a} \dot{\gamma}\right)+2 a k \gamma+k^{\frac{3}{2}} \alpha\right)}{4 a^3 \left(k r^2+4\right)}=0 \label{E[34]simple}
\end{eqnarray}

\noindent from which we can see that we only need to solve the two following ODEs in time for $\alpha(t)$, $\beta(t)$, and $\gamma(t)$.
\begin{eqnarray}
    6 a^2 b_3 \sqrt{k}  \dot{\gamma}+a k  \left(a \dot{\beta}+\dot{a} \beta+2 \dot{\alpha}\right)-2 \dot{a} k \alpha = 0 \label{simpleode12}\\
    -2 a^2 \left(a \ddot{\gamma}+3 \dot{a} \dot{\gamma}\right)+2 a k \gamma+k^{\frac{3}{2}} \alpha=0 \label{simpleode34}
\end{eqnarray}

There is, however, the issue that we have two equations for three unknown functions. This has occurred because even though we had three equation \eqref{presconstraint}-\eqref{E[34]v2} and three functions $\tilde{a},\tilde{m},\tilde{\chi}$, \eqref{E[34]v2} is satisfied just from the radial part of the simple ansatz \eqref{simpleansatz1}-\eqref{simpleansatz3} and hence we only have two equations that constrain the time parts of the functions. As in the FLRW case, we can solve this problem by asserting an equation of state. We will motivate a linear equation of state by studying the behaviour of $\tilde{\rho}$ and $\tilde{p}$ in the FLRW limit described in the previous section. We can solve for these functions by solving \eqref{E(11)}-\eqref{E(33)} and then substituting in the simple ansatz \eqref{simpleansatz1}-\eqref{simpleansatz3} to get

\begin{multline}
    \tilde{\rho}(t,r) = -3 r^2\frac{ 2k^2 \alpha + k^2 a \beta + 6 b_3 k^{\frac{3}{2}} a \gamma}{8 a^3 (4+kr^2)} \\ - 12\frac{ (-2k + 6b_3 k -4\dot{a}^2)\alpha +a((-k+3b_3k+2\dot{a}^2) \beta + 6b_3 \sqrt{k} \gamma + 4 \dot{a} \dot{\alpha} + 2 a \dot{a} \dot{\beta})}{8 a^3 (4+kr^2)} 
\end{multline}

\begin{multline}
    \tilde{p}(t,r) = r^2 \frac{k^2 \alpha +3 b_3 k^{\frac{3}{2}} a \gamma}{4a^3 (4+kr^2)} \\ + \frac{1}{4a^3 (4+kr^2)} \bigg[4\alpha ( k-3b_3k + 2\dot{a}^2 + 2a \ddot{a})-2a\big(6 b_3 \sqrt{k}\gamma +4 \dot{a} (\dot{\alpha} +2a\dot{\beta}) \\ +\beta(-2k + 3b_3k +2\dot{a}^2 + 4 a \ddot{a})+2a(2\ddot{\alpha}+a \ddot{\beta})\big) \bigg]
\end{multline}

\noindent For both of these expressions we can see that the second terms obey the FLRW limit (they are decreasing near $r=2$); however, the first terms do not. Since we have already asserted the boundary condition that our metric must obey the FLRW limit, we then must assert the corresponding constraint on the matter. Therefore, we will require the function $\tilde{p}-w\tilde{\rho}$ to obey the FLRW limit, which effectively is the condition that the first terms above must obey the equation of state in the FLRW limit for some arbitrary constant equation of state parameter $w$. Using the above expressions for $\tilde{\rho}$ and $\tilde{p}$ this condition becomes

\begin{equation}
    -2k^2(1+3w) \alpha - 3k^2wa\beta-6b_3 k^{\frac{3}{2}} (1+3w) a \gamma = 0 \label{eoseqn}
\end{equation}

\noindent which can be solved with

\begin{equation}
    \gamma(t) = -\frac{\sqrt{k}(2(1+3w)\alpha(t) +3wa(t)\beta(t))}{6b_3(1+3w)a(t)} \label{gammasoln}
\end{equation}

\noindent which is valid for $w \neq -1/3$ (if we had set $w=-1/3$ before solving for the equation of state condition, then \eqref{eoseqn} would have become $k^2 a \beta=0$, and since $k \neq 0$ for the bounce to occur, then the only way to satisfy the equation of state condition would have been $\beta=0$ which leads to $\tilde{m}=0$ due to \eqref{simpleansatz2} and this eliminates the central inhomogeneity). 

Substituting \eqref{gammasoln} back into \eqref{simpleode12} and \eqref{simpleode34} we have

\begin{equation}
    2ka \left( -\beta \dot{a} - \frac{a \dot{\beta}}{1+3w} \right) = 0 \label{simpleode12v2}
\end{equation}

\begin{multline}
    \frac{\sqrt{k}}{3 b_3(1+3w)}\bigg[(1+3 w)\alpha \left((3 b_3-2) k-2 \dot{a}^2-2a\ddot{a}\right)+a \bigg(a \left(3 a w \ddot{\beta}+(6 w+2) \ddot{\alpha} \right) \\ +\dot{a} \left(9 a w \dot{\beta} +(6 w+2) \dot{\alpha} \right)-3 k w \beta \bigg)\bigg]=0 \label{simpleode34v2}
\end{multline}

\noindent Immediately we can solve \eqref{simpleode12v2} with 

\begin{equation}
    \beta(t) = \frac{\beta_0}{a(t)^{1+3w}} \label{betasoln}
\end{equation}

\noindent where $\beta_0$ is the constant of integration. This leaves \eqref{simpleode34v2} in the form

\begin{equation}
    \frac{\sqrt{k}}{3b_3} \bigg( \alpha \left( (3b_3-2)k -2\dot{a}^2-2a\ddot{a} \right) - 3w\beta_0 a^{-3w}\left( \frac{k}{1+3w} + (1-3w)\dot{a}^2 + a \ddot{a} \right) + 2a \left( \dot{a} \dot{\alpha} + a \ddot{\alpha} \right) \bigg)=0 \label{simpleode34v3}
\end{equation}

\noindent Unfortunately, this equation cannot be solved exactly in closed form for $\alpha(t)$ (without expressing it in terms of an integral). We can solve it by expanding $\alpha(t)$ and $a(t)$ in a series in $t$ up to $t^4$ order and solving for the coefficients of $\alpha(t)$ up to $t^2$ order of the differential equation. Since the differential equation is second order in time, this will give us the most general series solution up to $t^2$ order. Then once again substituting \eqref{ascsolnpert} with $\alpha(t) = \alpha_0 + \alpha_1 t^1 + \alpha_2 t^2 + \alpha_3 t^3 + \alpha_4 t^4$ we obtain

\begin{multline}
    \bigg[ \frac{\sqrt{k}}{3 b_3} \left(\frac{1}{3} \Lambda  (a_-^2-a_+^2) a_+^{-3 w} \left(2 \alpha_0 a_+^{3 w}+3 \beta_0 w\right)-\frac{3 \beta_0 k w a_+^{-3 w}}{1+3w}+4 \alpha_2 a_+^2+\alpha_0 (3 b_3-2) k\right) \bigg] \\ + t \bigg[ \frac{\sqrt{k} \left(12 \alpha_3 a_+^2+\alpha_1 (3 b_3-2) k\right)}{3 b_3} \bigg] \\ + t^2 \bigg[ \frac{\sqrt{k}}{54 b_3} \bigg(\frac{3 \beta_0 \Lambda  w (a_-^2-a_+^2) a_+^{-3 w-2} \left(\Lambda  (1+3w) \left(9 a_-^2 w+a_+^2 (4-9 w)\right)-9 k w\right)}{1+3w} \\ +4 \Lambda  (a_-^2-a_+^2) (2 \alpha_0 \Lambda -9 \alpha_2)+432 \alpha_4 a_+^2+18 \alpha_2 (3 b_3-2) k \bigg) \bigg] + \mathcal{O}(t^3) = 0
\end{multline}

\noindent Solving for the $\alpha_i$'s, we have that

\begin{multline}
    \alpha(t) = \alpha_0 + \alpha_1 t \\ + t^2 \bigg[ \frac{1}{12} \left(2 \alpha_0 \Lambda -\frac{3 \beta_0 w a_+^{-3 w-2} \left(a_-^2 \Lambda  (1+3w)-3 k\right)}{1+3w}+\frac{\alpha_0 \left((6-9 b_3) k-2 a_-^2 \Lambda \right)}{a_+^2}+3 \beta_0 \Lambda  w a_+^{-3 w}\right) \bigg] \label{alphasoln}
\end{multline}

\noindent where $\alpha_0$ and $\alpha_1$ represent the two constants that come from solving a second order ODE. Putting together  \eqref{gammasoln}, \eqref{betasoln}, and \eqref{alphasoln} we have a solution to the $\epsilon^1$ order corrections to FLRW coming from the general McVittie metric in NGR. To make physical predictions from this solution, we must constrain the free constants by asserting some initial conditions.

\subsubsection{Initial Conditions and GR Limit}

We have three arbitrary constants to consider; $\alpha_0, \alpha_1$, and $\beta_0$. We can use the condition that we have a well behaved GR limit when $b_3 \to 0$ to constrain these constants. We can see the issue in the generic case by looking at \eqref{gammasoln}

\begin{equation}
    \gamma(t) = -\frac{\sqrt{k}(2(1+3w)\alpha(t) +3wa\beta(t))}{6b_3(1+3w)a(t)} 
\end{equation}

\noindent At first glance, it seems $\gamma$ diverges as $b_3 \to 0$; however, the functions $\alpha$, $\beta$, and $a(t)$ also contains $b_3$'s. Substituting the solution for $\alpha$, $\beta$ and for the scale factor $a(t)$ and expanding up to $t^2$ we have that

\begin{multline}
    \gamma(t) = \bigg[ -\frac{\sqrt{k} \left(3 \beta_0 w a_+^{-3 w}+\alpha_0 (6 w+2)\right)}{6 a_+ b_3 (3 w+1)} \bigg] + \bigg[ -\frac{\alpha_1 \sqrt{k}}{3 a_+ b_3} \bigg]t \\ + \bigg[ \frac{k^{3/2} a_+^{-3 (w+1)} \left(\alpha_0 (3 b_3-2) (3 w+1) a_+^{3 w}-3 \beta_0 w\right)}{12 (3 b_3 w+b_3)} \bigg]t^2 + \mathcal{O}(t^3) \label{gammapreIC}
\end{multline}

\noindent Where $a_\pm$ contains $b_3$ via \eqref{apm}. We want to expand the coefficients of each order of $t$ from \eqref{gammapreIC} to leading order in $b_3$ as $b_3 \to 0$ and find the conditions on $\alpha_0$ and $\alpha_1$ such that these terms are finite in this limit. These coefficients become

\begin{multline}
    t^0 : -\frac{1}{3 \sqrt{2} b_3 (3 w+1)} \bigg[ \sqrt{k \Lambda } \left(\sqrt{9 k^2-4 \Lambda  \rho_0}+3 k\right)^{-\frac{3 w}{2}-\frac{1}{2}} \left(2 \alpha_0 (3 w+1) \left(\sqrt{9 k^2-4 \Lambda  \rho_0}+3 k\right)^{\frac{3 w}{2}} \right. \\ \left. +3 \beta_0 2^{\frac{3 w}{2}} w \Lambda ^{\frac{3 w}{2}}\right) \bigg] + \mathcal{O}(1) 
\end{multline}

\begin{equation}
    t^1 : -\frac{\sqrt{2} \alpha_1 \sqrt{k \Lambda }}{3 b_3 \sqrt{\sqrt{9 k^2-4 \Lambda  \rho_0}+3 k}}+ \mathcal{O}(1) 
\end{equation}

\begin{multline}
    t^2 : -\frac{1}{\sqrt{2} b_3 (3 w+1)} \bigg[ k^{3/2} 3^{-\frac{3 w}{2}-\frac{5}{2}} \exp \left(\frac{3}{2} (w+1) \left(\ln (\Lambda )-\log \left(\frac{1}{3} \sqrt{9 k^2-4 \Lambda  \rho_0}+k\right)\right)\right) \\ \left(2 \alpha_0 (3 w+1) \left(\frac{\sqrt{9 k^2-4 \Lambda  \rho_0}+3 k}{\Lambda }\right)^{\frac{3 w}{2}}+3 \beta_0 2^{\frac{3 w}{2}} w\right) \bigg] + \mathcal{O}(1)
\end{multline}

\noindent The terms that are of of order $\mathcal{O}(1/b_3)$ will diverge as $b_3 \to 0$. Hence, to obtain regular functions in this limit, we need to choose $\alpha_0$ and $\alpha_1$ such that the above terms of order $\mathcal{O}(1/b_3)$ are zero. These conditions are satisfied by

\begin{eqnarray}
    \alpha_0 &=& -\frac{3 \beta_0 2^{\frac{3 w}{2}-1} w \left(\frac{\sqrt{9 k^2-4 \Lambda  \rho_0}+3 k}{\Lambda }\right)^{-\frac{3 w}{2}}}{3 w+1} \\
    \alpha_1 &=& 0
\end{eqnarray}

\noindent Since we have written our solution in terms of $a_\pm$ we can write $\alpha_0$ as

\begin{equation}
    \alpha_0 = -\frac{3 \beta_0 2^{\frac{3 w}{2}-1} w \left(\frac{\sqrt{\Lambda ^2 \left(a_-^2-a_+^2\right)^2+\frac{27}{4} b_3 (4-3 b_3) k^2}+3 k}{\Lambda }\right)^{-\frac{3 w}{2}}}{3 w+1} 
\end{equation}

\noindent We then obtain $\alpha(t)$ up to $t^2$ order:

\begin{multline}
    \alpha(t) = \bigg[ -\frac{3 \beta_0 2^{\frac{3 w}{2}-1} w}{3 w+1} \left(\frac{1}{\Lambda }\sqrt{\Lambda ^2 \left(a_-^2-a_+^2\right)^2+\frac{27}{4} b_3 (4-3 b_3) k^2}+3 k\right)^{-\frac{3 w}{2}} \bigg] \\ + \bigg[ \frac{\beta_0 w}{8 a_+^2 (3 w+1)} \bigg(  8^w (-2 \Lambda  (a_+^2-a_-^2) \\  +(9 b_3-6) k) \left(\frac{1}{\Lambda }\sqrt{4 \Lambda ^2 \left(a_-^2-a_+^2\right)^2+27 b_3 (4-3 b_3) k^2}+6 k\right)^{-\frac{3 w}{2}}  \\  +a_+^{-3 w} (6 k+2 \Lambda  (3 w+1) (a_+^2-a_-^2) ) \bigg) \bigg] t^2 + \mathcal{O}(t^3) \label{finalalpha}
\end{multline}

\noindent Hence, $\tilde{\chi}$ is regular in the GR limit. Since $\tilde{\psi}$ is defined via $\tilde{\chi}$ in our solution \eqref{solvetpsi} then $\tilde{\psi}$ is regular in this limit as well.

\subsubsection{Conservation Equations}

For completeness, we show that the conservation equations up to $\epsilon^1$ order are satisfied. There are two conservation equations corresponding to $\accentset{\circ}{\nabla}_\mu \Theta \indices{^\mu _1}=0$ and $\accentset{\circ}{\nabla}_\mu \Theta \indices{^\mu _2}=0$ where $\accentset{\circ}{\nabla}$ is the covariant derivative with respect to the Levi-Civita connection. Note that the $\epsilon^0$ order conservation equations are already satisfied by the choice of the ansatz \eqref{asymansatzA}-\eqref{asymansatzchi}. After substituting in the simple ansatz \eqref{simpleansatz1}-\eqref{simpleansatz3}, the two conservation equations become

\begin{multline}
    \frac{1}{8 a^6 \left(k r^2+4\right)} \bigg[ 2\alpha \dot{a}\left(-16 \rho_0+3a^2\left(k \left(-12 b_3+k r^2+4\right)+8 \dot{a}^2-8a\ddot{a}\right)\right) \\ +a\left(32 \rho_0 \dot{\alpha}+a\left(16 \rho_0 \dot{\beta}+3a\left(\dot{a} k \left(4-k r^2\right) \beta -\left(8 \dot{a}^2+k \left(-12 b_3+k r^2+4\right)\right) \left(a \dot{\beta}+2 \dot{\alpha}\right) \right. \right. \right. \\ \left. \left. \left. -6 a b_3 \sqrt{k} \left(k r^2-4\right) \dot{\gamma}+8aa \dot{\beta}+2 \dot{\alpha}\ddot{a}\right)\right)\right) \bigg]=0 \label{cons1}
\end{multline}

\begin{multline}
    \frac{kr}{48 a^6}\bigg[4 \rho_0 \beta +3a\left(\alpha \left(6 b_3 k-4 \dot{a}^2-4a\ddot{a}\right)+a\left(12 b_3 \sqrt{k} \gamma +4 \dot{a} \left(2 a \dot{\beta}+\dot{\alpha}\right) \right. \right. \\ \left. \left. +\beta \left(-2 k+3 b_3 k+2 \dot{a}^2+4a\ddot{a}\right)+2 a \left(a \ddot{\beta}+2 \ddot{\alpha}\right)\right)\right) \bigg] =0 \label{cons2}
\end{multline}

\noindent Substituting the solutions for $\beta(t)$ and $\gamma(t)$ (\eqref{betasoln}, \eqref{gammasoln}) into the above conservation equations, we obtain

\begin{multline}
    \frac{a^{-3 (w+2)}}{2 \left(k r^2+4\right)} \left( \dot{a} \left(2 a^{3 w} \alpha +\beta_0+3 \beta_0 w\right) -2 a^{1+3w} \dot{\alpha} \right) \left(-4 \rho_0+3a^2\left((2-3 b_3) k+2 \dot{a}^2-2a\ddot{a}\right)\right) \\ =0 \label{cons1v2}
\end{multline}

\begin{multline}
    \frac{ka^{-3 w-7}r}{48 (1+3w)}\bigg[ 4 \beta_0 \rho_0 (1+3w)+3a^2\bigg( \beta_0\left(k (9 b_3 w+3 b_3-12 w-2)+2 \dot{a}^2 \left(27 w^3-6 w-1\right) \right. \\ \left. +2(1-9 w^2)a\ddot{a}\right)  +2(1+3w)a^{3 w}\left(\alpha \left(-2 k+3 b_3 k-2 \dot{a}^2-2a\ddot{a}\right)+2 a \left(a \ddot{\alpha}+\dot{a} \dot{\alpha}\right)\right) \bigg) \bigg] =0 \label{cons2v2}
\end{multline}

\noindent Note that \eqref{cons1v2} is automatically satisfied when we substitute \eqref{flrwsoln} for $a(t)$ since  $-4 \rho_0+3a^2\big((2-3 b_3) k $ \\ $+2 \dot{a}^2-2a\ddot{a}\big)=0$. However, \eqref{cons2v2} is not automatically satisfied, because we have not yet used the condition that \eqref{simpleode34v3} is solved for $\alpha(t)$. Without exactly solving for $\alpha(t)$, we can substitute $\ddot{\alpha}$ from \eqref{simpleode34v3} into \eqref{cons2v2} and substitute \eqref{flrwsoln}, in which case these conservation equations are satisfied.

\subsection{Summary of Perturbative Solution}

We have found an asymptotic solution that is valid near $t=0$, which is when the bounce occurs and is the event that we are concerned with. We can write the full solution to the $\epsilon^1$ order functions (tilde functions) as follows:

\begin{eqnarray}
    \tilde{a}(t,r) &=& \frac{1}{4+ kr^2} \alpha(t)  \\
    \tilde{m}(t,r) &=& \frac{r}{(4+kr^2)^{\frac{3}{2}}} \frac{\beta_0}{a(t)^{1+3w}}
\end{eqnarray}

\begin{multline}
    \tilde{\rho}(t,r) = -\frac{3k^2\beta_0 r^2}{8(4+kr^2)(1+3w)a(t)^{3(1+w)}} \\ + 12 \frac{k\left( -1-6w+3b_3(1+3w) \right)\beta_0+2(1+3w) \left( -3w\beta_0 \dot{a}^2 + a^{3w} \left( \alpha \left( (-2+3b_3)k-2\dot{a}^2 \right) + 2a \dot{a} \dot{\alpha} \right) \right)}{8(4+kr^2)(1+3w)a(t)^{3(1+w)}} 
\end{multline}

\begin{multline}
    \tilde{p}(t,r) =  -\frac{3k^2\beta_0 w r^2}{8(4+kr^2)(1+3w)a(t)^{3(1+w)}} \\ -\frac{4}{8(4+kr^2)(1+3w)a(t)^{3(1+w)}} \bigg[ k\left( -2+3b_3+9(-1+b_3)w \right)\beta_0-2\beta_0\dot{a}^2+6w \left( -2+9w^2 \right)\beta_0 \dot{a}^2 \\ +2\left( 1-9w^2 \right)\beta_0 a \ddot{a} + 2(1+3w) a^{3w}\left( \alpha \left( -2k +3b_3k -2 \dot{a}^2-2a \ddot{a} \right) +2a \left( \dot{a} \dot{\alpha} + a \ddot{\alpha} \right) \right) \bigg] 
\end{multline}

\begin{eqnarray}
    \tilde{\psi}(t,r) &=& - \frac{r\left( a^{-1-3w} \left( 3w\beta_0+ 2 a^{3w} \alpha \right) \dot{a}-2\dot{\alpha} \right)}{6b_3(4+kr^2)} \label{psifinal} \\
    \tilde{\chi}(t,r) &=& -\frac{r}{4 + kr^2} \frac{\sqrt{k} \left(2(1+3w)\alpha(t) + \frac{3wa\beta_0}{a(t)^{1+3w}} \right)}{6b_3(1+3w)a(t)} \label{chifinal}
\end{eqnarray}

\noindent where $a(t)$ is given by \eqref{ascsolnpert} and $\alpha(t)$ is given by \eqref{finalalpha}. We can also see, after we substitute the solution of $\alpha(t)$ into the above functions representing the perturbations, that in all of these expressions we can factor out a $\beta_0$. This then implies that when we set the artificial perturbative parameter $\epsilon = 1$, the true ``small" factor that represents the perturbations is $\beta_0$.

\subsection{Local Horizon}

We can locate the horizon locally by solving the necessary condition $\theta_{(l)}\theta_{(n)}=0$ for $r=r_h(t)$  where $\theta_{(l)}$ and $\theta_{(n)}$ are the outgoing/ingoing null expansion scalars, respectively \cite{perez}. In the teleparallel framework, these expansions take the form

\begin{equation}
    \theta_{(\ell)}
    = \nabla_{\mu}\ell^{\mu}
    + K^{\sigma\mu}{}_{\mu}\,\ell_{\sigma},
    \qquad
    \theta_{( n)}
    = \nabla_{\mu}n^{\mu}
    + K^{\sigma\mu}{}_{\mu}\,n_{\sigma},
\end{equation}

\noindent where $ \nabla_{\mu} $ denotes the covariant derivative with respect to the teleparallel connection \eqref{tgconnection}, $K_{\sigma\mu\nu} = T_{[\mu\sigma]\nu} + \tfrac{1}{2} T_{\nu\sigma\mu}$ is the contortion tensor, and $l^\mu, n^\nu$ are the outgoing and ingoing null vectors, respectively, given by

\begin{eqnarray}
    l^\mu &=& \left( \frac{1}{\sqrt{2}A_1(t,r)}, \frac{1}{\sqrt{2}A_2(t,r)},0,0 \right) \\ 
    n^\mu &=& \left( \frac{1}{\sqrt{2}A_1(t,r)}, -\frac{1}{\sqrt{2}A_2(t,r)},0,0 \right)
\end{eqnarray}

\noindent We calculate the location of the horizon by first expanding $\theta_{(l)}\theta_{(n)}$ in terms of $\epsilon$:

\begin{equation}
    \theta_{(l)}\theta_{(n)} = f(t,r) + \epsilon g(t,r) + \mathcal{O}(\epsilon^2)
\end{equation}

\noindent For some functions $f,g$. Since there are two independent equations to solve due to the asymptotic expansion, namely $f(t,r)=0$ and $g(t,r)=0$, we will need to also expand the horizon radius in terms of $\epsilon$:

\begin{equation}
    r_h(t) = r_0(t) + \epsilon r_1(t) \label{hradius}
\end{equation}

\noindent Substituting this into $\theta_{(l)}\theta_{(n)}$ then expanding in $\epsilon$ once again, we find that

\begin{equation}
    \theta_{(l)}\theta_{(n)} = f(t,r_0(t)) + \epsilon \left(f'(t,r_0(t)) r_1(t) + g(t,r_0(t) \right) + \mathcal{O}(\epsilon^2)
\end{equation}

\noindent where $f'(t,r_0(t)) \equiv \partial_r f(t,r)|_{r=r_0(t)}$. This means that to solve $\theta_{(l)}\theta_{(n)}=0$ to $\epsilon^0$ order, we simply need to solve for $r_0(t)$ such that $f(t,r_0(t))=0$, and then to the $\epsilon^1$ order we require

\begin{equation}
    r_1(t) = - \frac{g(t,r_0(t)}{f'(t,r_0(t))}
\end{equation}

\noindent Consequently, \eqref{hradius} will give the location of the horizon locally as a function of time up to $\epsilon^1$ order. Due to the strange behaviour of the radial coordinate in isotropic coordinates, we will use the areal radius coordinate, $R$, to express the location of the horizon as is often done in the literature \cite{perez}.

To transform from $r$ to $R$, we can write \eqref{arealrad} as $r = \frac{2}{R}(a \pm \sqrt{a^2 - R^2})$. There are two branches given by the choice of $\pm$, one corresponding to the $r \in (0,2)$ region while the other to the $(2,\infty)$ region. To differentiate between the two we use the fact that $r = 0$ and $r = \infty$ both correspond to $R=0$. To leading order the values of the above relation between $r$ and $R$ as $R \to 0$ yield

\begin{equation}
    r_\pm \sim \frac{2}{R}(a\pm a)
\end{equation}

\noindent The $+$ branch gives $r_+ \sim \frac{4}{R}a$, which tends to infinity as $R \to 0$, and the $-$ branch gives $r_- \sim 0$. In other words, $r_+ \in (2,\infty)$ and $r_- \in (0,2)$. Therefore, to stay within the $(0,2)$ region we choose the transformation given by the $-$ branch $r = \frac{2}{R}(a - \sqrt{a^2 - R^2})$. The reason this is important is because if had chosen the other branch, the null vectors $l^\mu$ and $n^\mu$ would swap their roles of being outgoing or ingoing.

The general condition to find the location of the local horizon is $\theta_{(l)}=0$ or $\theta_{(n)}=0$, which is the case will depend on the sign of $\dot{a}$. Thus, we calculate both of the expansion scalars up to $\epsilon^1$ order

\begin{multline}
    \theta_{(l)} = \bigg[\frac{\sqrt{2}(\sqrt{a^2-k R^2}+R \dot{a})}{ R a} \bigg] \\ + \epsilon \bigg[ \frac{k R \left(4 a^3+(-4 a^2 +kR^2) \sqrt{a^2-kR^2}-3 a kR^2\right) \left(a \left(2R  \dot{\alpha}-2 \alpha  +\dot{a} \beta R\right)-2 \dot{a}R \alpha  +a^2 \left(R\dot{\beta}-\beta \right)\right)}{8 \sqrt{2} a^3 \left(a-\sqrt{a^2-kR^2}\right)^4 } \bigg]
\end{multline}

\begin{multline}
    \theta_{(n)} = \bigg[\frac{\sqrt{2}(-\sqrt{a^2-k R^2}+R \dot{a})}{ R a} \bigg] \\ + \epsilon \bigg[ \frac{kR \left(4 a^3+(-4 a^2 +kR^2) \sqrt{a^2-kR^2}-3 a kR^2\right) \left(a \left(2R  \dot{\alpha}+2 \alpha  +\dot{a} \beta R\right)-2 \dot{a}R \alpha +a^2 \left(R\dot{\beta}+\beta \right)\right)}{8 \sqrt{2} a^3 \left(a-\sqrt{a^2-kR^2}\right)^4 } \bigg]
\end{multline}

\noindent Notice that the FLRW ($\epsilon=0$) horizon is given by the well known expression

\begin{equation}
    R_0(t) = \frac{1}{\sqrt{\frac{\dot{a}^2}{a^2}+\frac{k}{a^2}}}
\end{equation}

\noindent In general, the location of the horizon, in terms of the areal radial coordinate $R$, is given up to $\epsilon^1$ order by

\begin{multline}
    0 =\theta_{(l)} \theta_{(n)} = \bigg[\frac{2 \left(\dot{a}^2+k\right)}{a^2}-\frac{2}{R^2} \bigg] \\ + \epsilon \bigg[\frac{1}{4 a^4 R^2} \bigg(a^4 \beta +a^2 \left(\sqrt{a^2-k R^2} \left(\dot{a} R^2 \dot{\beta}+2 \alpha \right)+R^2 \left(\left(\dot{a}^2-k\right) \beta +2 \dot{a} \dot{\alpha}\right)\right)\\+a R^2 \left(\dot{a} \sqrt{a^2-k R^2} \left(\dot{a} \beta +2 \dot{\alpha}\right)-2 \left(\dot{a}^2+k\right) \alpha \right)-2 \dot{a}^2 R^2 \alpha  \sqrt{a^2-k R^2}\\+a^3 \left(\beta  \sqrt{a^2-k R^2}+\dot{a} R^2 \dot{\beta}+2 \alpha \right) \bigg) \bigg] 
\end{multline}

\noindent Then, using the procedure above, we can calculate the radius of the local horizon $R_h(t) = R_0(t) + \epsilon R_1 (t)$ by substituting the solution for $\alpha,\beta,\gamma$ \eqref{gammasoln},\eqref{betasoln},\eqref{alphasoln} with the initial conditions and expand both $R_0(t)$ and $R_1(t)$ in terms of time to get

\begin{equation}
    R_0(t) = \bigg[a_+\bigg] + \bigg[ \frac{\Lambda  \left(a_+^2-a_-^2\right) \left(3- \Lambda(a_+^2 -a_-^2 )  \right)}{18 a_+ } \bigg] t^2 + \mathcal{O}(t^3) \label{hor0}
\end{equation}

\begin{multline}
    R_1(t) = \bigg[\frac{1}{48 a_+} \beta_0 \Lambda  (a_-^2-a_+^2) \left(a_+^{-3 w}-\frac{3\ 8^w w}{3 w+1} \left(\frac{1}{\Lambda }\sqrt{4 \Lambda ^2 \left(a_-^2-a_+^2\right)^2+27 b_3 (4-3 b_3)}+6 \right)^{-\frac{3 w}{2}}\right) \bigg]t \\ + \bigg[ \frac{\beta_0 \Lambda  (a_-^2-a_+^2)}{288 a_+^2 (3 w+1)} \bigg(3\ 8^w w (2 \Lambda  (a_-^2-a_+^2)+9 b_3-6) \left(\frac{1}{\Lambda }\sqrt{4 \Lambda ^2 \left(a_-^2-a_+^2\right)^2+27 b_3 (4-3 b_3)}+6\right)^{-\frac{3 w}{2}} \\ +a_+^{-3 w} (18 w-2 \Lambda  (3 w+1) (a_-^2-a_+^2))\bigg) \bigg]t^2 + \mathcal{O}(t^3) \label{hor1}
\end{multline}

\section{Discussion}

Starting from a generalized McVittie metric \eqref{genmetric}, we have employed a perturbative and local approach to model a bouncing cosmological background with a central inhomogeneity in a one parameter NGR theory described by \eqref{action}. This perturbed solution has leading order behaviour described by NGR FLRW and the central inhomogeneity comes in at higher orders. Previously, black hole persistence has not been studied within modified theories of gravity within which a bounce is perhaps most justified. This is possibly due to the complexity of the equations of a McVittie spacetime combined with the equations of a modified theory. Here we have presented one method to study the effects of a central inhomogeneity during a bounce in a cosmological model. We have used NGR as an example of a modified theory of gravity; however, this method is suited to other theories such as scalar-tensor theories. 

It is important to note that the approach taken here utilizing the generalized McVittie geometry and affecting the bounce within NGR is not the only possible approach to study a black hole embedded in a bouncing universe. In particular, there are a number of other classical bounce mechanisms and in the future we intend to study the problem within scalar tensor theories of gravity. However, perhaps the most sensible approach is via numerical methods \cite{corman}.

We can analyze how the perturbations affect the leading order terms which will depend on the choice of constants $k,\beta_0, b_3$ and $w$, some of which are contained within $a_\pm$ via \eqref{apm} within which $\Lambda$ and $\rho_0$ can, in principle, be constrained by observation. The focus is on the sign of the perturbative terms, which is sufficient to determine a qualitative understanding. We only consider the $k=+1$ case as it is required for our perturbative solution to model a central inhomogeneity and $b_3 >0$ to satisfy the ghost free condition of the theory \cite{diego}. This leaves us with $\beta_0, b_3>0$ and $w$ to consider. We also discover that $\beta_0$ is the true perturbative parameter which allows us to assume $\beta_0 \ll 1$.

Starting with the leading order behaviour, NGR modifies the GR FLRW solution by simply re-normalizing the curvature term in the Friedman equation. The local horizon \eqref{hor0} then is also re-normalized by NGR through a change in the values of $a_\pm$, where in GR we have $B=1$. 

Looking at the perturbation to the scale factor coming from $\tilde{a}$, the solution found for $\alpha(t)$ \eqref{finalalpha} shows that the perturbation changes the minimum of the bounce by modifying the $t^0$ order term (as well as the $t^2$ order term). But it does not introduce a term at the $t^1$ order, hence near $t=0$ the bounce remains symmetric. Different choices of the above mentioned free constants will determine the sign of the perturbations and hence whether the minimum of the scale factor is increased or decreased from the $t^0$ order and whether the scale factor's rate of change is increased or decreased from the $t^2$ order terms. For example, for $k=+1$, the sign of the factor $-\frac{\beta_0 w}{1+3w}$ will determine the sign of the $t^0$ order term of the perturbation to the scale factor. The sign of the $t^2$ term is more complicated but depends on the same constants.

Considering the local horizon, $R_0(t)$ \eqref{hor0} shows that the horizon experiences a bounce of the form $R_0 \sim c_0 + c_2 t^2$, where the sign of the $t^2$ order term is positive ($c_2 > 0$) hence it contracts and expands with the bounce. Then the perturbation $R_1(t)$ \eqref{hor1} does not affect the minimum of the horizon $R_0 \sim a_+$, rather it affects the higher order terms, the sign of which only speeds up or slows down the dynamics of the horizon. One thing to note is that the leading order horizon $R_0$ is symmetric across $t=0$, while the perturbation $R_1$ disrupts this symmetry due to the $t^1$ order term.

\subsection*{Acknowledgements}

D.F.L. and B.Y. acknowledges support from the Department of Mathematics and Statistics at Dalhousie University, Canada. A.A.C. is supported by the Natural Sciences and Engineering Research Council of Canada (NSERC).


\begin{thebibliography}{10}

\bibitem{novello}
M. Novello and S. Bergliaffa, Phys. Rep. {\bf 463}, 127 (2008) [arXiv:0802.1634].

\bibitem{Tolman31}
R.C. Tolman, Phys. Rev. {\bf 38}, 1758 (1931).

\bibitem{Lemaitre33}
G. Lemaitre, Ann. Soc. Sci. Brux. {\bf 53}, 51 (1933).

\bibitem{Lemaitre:1933gd}
G.~Lemaitre, \newblock Gen. Rel. Grav. {\bf 29}, 641 (1997).

\bibitem{lemaitre}
G. Lemaitre, Mon. Not. R. Astron. Soc. {\bf 91}, 490-501 (1931).

\bibitem{string}
N. Turok, M. Perry and P. J. Steinhardt, Phys. Rev. D {\bf{70}}, 106004 (2004).

\bibitem{BHScorr}
G. Veneziano, Europhys. Lett. {\bf 2}, 133 (1986).

\bibitem{sing}
M. Bojowald, Phys.\ Rev.\ Lett. {\bf 86}, 5227 (2001); M. Bojowald, Phys.\ Rev.\ Lett. {\bf 95}, 091302 (2005); A. Ashtekar, T. Pawlowski and P. Singh, Phys.\ Rev.\ D {\bf 74}, 084003 (2006).

\bibitem{ashtekar}
A. Ashtekar, in {\it Einstein and the Changing Worldviews of Physics}, ed. C. Lehner, J. Renn and M. Schemmel, Springer, Berlin (2012).

\bibitem{quintin}
J. Quintin and R. Brandenberger, JCAP, {\bf 11}, 029 (2016) [arXiv:1609.02556]; Y-F. Cai, R. Brandenburger and P. Peter (2013) [arXiv:1301.4703]; R. -G. Cai and A. Wang, Phys. Rev. D {\bf 73}, 063005 (2006); Y-F. Cai, W. Xue, R. Brandenburger and Zhang, (2009) [arXiv:0903.4938].

\bibitem{Battefeld_2015}
D. Battefeld and P. Peter, \newblock Physics Reports {\bf 571}, 1 (2015) [arXiv:1406.2790].

\bibitem{brand}
R.~Brandenberger and P.~Peter, Found. Phys. \textbf{47}, 6, 797-850 (2017) [arXiv:1603.05834].

\bibitem{cyclic}
N. Itzhaki, U. Peleg and P J. Steinhardt, (2026) [arXiv:2508.09745]; .-L. Lehners, Class. Quant. Grav. {\bf 28}, 204004 (2011); A. Ijjas and P. J. Steinhardt, Phys. Lett. B {\bf 795}, 666–672 (2019); P. J. Steinhardt and N. Turok, Science {\bf 296}, 1436–1439 (2002).

\bibitem{Martin:2003sf}
J.~Martin and P.~Peter, \newblock Phys. Rev. D {\bf 68}, 103517 (2003).

\bibitem{Galkina_2019}
O.~Galkina, J.~C. Fabris, F.~T. Falciano, and N.~Pinto-Neto, \newblock JETP Lett. {\bf 110}, 523–528 (2019) [arXiv:1908.04258].

\bibitem{ijjas}
A. Ijjas and P. J. Steinhardt (2021) [arXiv:2108.07107] ;{\em{ibid.}} Phys. Lett. B {\bf 764}, 289 (2017) [arXiv:1609.01253]; {\em{ibid.}} Phys. Rev. Lett. {\bf 117}, 121304 (2016) [arXiv:1606.08880]

\bibitem{lib}
M. Libanov, S. Mironov and V. Rabakov, (2016) [arXiv:1605.05992]; T. Kobayashi, (2016) [aXiv:1606.05831].

\bibitem{Dobre:2017pnt}
D.~A.~Dobre, A.~V.~Frolov, J.~T.~G{\'a}lvez Ghersi, S.~Ramazanov and A.~Vikman,
JCAP \textbf{03}, 020 (2018)
[arXiv:1712.10272].

\bibitem{Nicolis_2009Galileon}
A. Nicolis, R. Rattazzi, and E. Trincherini, \newblock Phys. Rev. D \textbf{79}, 064036 (2009) [arXiv:0811.2197].

\bibitem{deRham:2010ik}
C.~de~Rham and G.~Gabadadze, \newblock Phys. Rev. D {\bf 82}, 044020 (2010) [arXiv:1007.0443].

\bibitem{deRham:2010kj}
C.~de~Rham, G.~Gabadadze, and A.~J. Tolley, Phys. Rev. Lett. {\bf 106}, 231101 (2011) [arXiv:1011.1232].

\bibitem{Creminelli:2010ba}
P.~Creminelli, A.~Nicolis, and E.~Trincherini, \newblock JCAP {\bf 1011}, 021 (2010) [arXiv:1007.0027].

\bibitem{Creminelli:2012my}
P.~Creminelli, K.~Hinterbichler, J.~Khoury, A.~Nicolis, and E.~Trincherini, \newblock JHEP {\bf 1302}, 006 (2013).

\bibitem{Qiu:2011cy}
T.~Qiu, J.~Evslin, Y.-F. Cai, M.~Li, and X.~Zhang, \newblock JCAP {\bf 1110}, 036 (2011) [arXiv:1108.0593].

\bibitem{Osipov:2013ssa}
M.~Osipov and V.~Rubakov, \newblock JCAP {\bf 1311}, 031 (2013) [arXiv:1303.1221].

\bibitem{Easson:2011zy}
D.~A. Easson, I.~Sawicki, and A.~Vikman, \newblock JCAP {\bf 11}, 021 (2011) [arXiv:1109.1047].

\bibitem{Feng:2004ad}
B.~Feng, X.L. Wang, X.M. Zhang, \newblock Phys. Lett. B {\bf 607}, 35, (2005) [arXiv:astro-ph/0404224].

\bibitem{Feng:2004ff}
B.~Feng, M.~Li, Y.S. Piao,X. Zhang , \newblock Phys. Lett. B {\bf 634}, 101, (2006).

\bibitem{Biswas:2010zk}
T.~Biswas, T.~Koivisto, and A.~Mazumdar, \newblock JCAP {\bf 1011}, 008 (2010) [arXiv:1005.0590].

\bibitem{Cai:2012ag}
Y.-F. Cai, C.~Gao, and E.~N. Saridakis, \newblock JCAP {\bf 1210}, 048 (2012).

\bibitem{Bacalhau_2018}
A.~P. Bacalhau, N.~Pinto-Neto, and S.~D.~P. Vitenti, \newblock Phys. Rev. D {\bf 97}, 083517 (2018) [arXiv:1706.08830].

\bibitem{Gasperini:2002bn}
M.~Gasperini and G.~Veneziano, \newblock Phys. Rept. {\bf 373}, 1 (2003) [arXiv:hep-th/0207130].

\bibitem{Khoury:2001wf}
J.~Khoury, B.~A. Ovrut, P.~J. Steinhardt, and N.~Turok, \newblock Phys. Rev. D {\bf 64}, 123522 (2001) [arXiv:hep-th/0103239].

\bibitem{Steinhardt:2001st}
P.~J. Steinhardt and N.~Turok, \newblock Phys. Rev. D {\bf 65}, 126003 (2002) [arXiv:hep-th/0111098].

\bibitem{Khoury:2001bz}
J.~Khoury, B.~A. Ovrut, N.~Seiberg, P.~J. Steinhardt, and N.~Turok, \newblock Phys. Rev. D {\bf 65}, 086007 (2002) [arXiv:hep-th/0108187].

\bibitem{Horava:2009uw}
P.~Horava, \newblock Phys. Rev. D {\bf 79}, 084008 (2009) [arXiv:0901.3775].

\bibitem{Horava:2009if}
P.~Horava, \newblock Phys. Rev. Lett. {\bf 102}, 161301 (2009) [arXiv:0902.3657].

\bibitem{Kiritsis:2009sh}
E.~Kiritsis and G.~Kofinas, \newblock Nucl. Phys. B {\bf 821}, 467 (2009) [arXiv:0904.1334].

\bibitem{Brandenberger:2009yt}
R.~Brandenberger, \newblock Phys. Rev. D {\bf 80}, 043516 (2009) [arXiv:0905.1514].

\bibitem{DeFelice:2010aj}
A.~D. Felice and S.~Tsujikawa, \newblock Living Rev.Rel. {\bf 13}, 3 (2010) [arXiv:1002.4928].

\bibitem{Bamba:2013fha}
K. Bamba, A. N. Makarenko, A. N. Myagky, S. Nojiri, and S. D. Odintsov (2013) [arXiv:1309.3748].

\bibitem{Bamba:2014mya}
K. Bamba, A. N. Makarenko, A. N. Myagky, and S. D. Odintsov (2014) [arXiv:1403.3242].

\bibitem{Cai:2009in}
Y.F. Cai and E.~N. Saridakis, \newblock JCAP {\bf 10}, 020 (2009) [arXiv:0906.1789].

\bibitem{Kibble:1961ba}
T.~Kibble, \newblock J.Math.Phys. {\bf 2}, 212 (1961).

\bibitem{Sciama:1964wt}
D.~W. Sciama, \newblock Rev. Mod. Phys. {\bf 36}, 463 (1964).

\bibitem{Cai:2011tc}
Y.F. Cai, S.-H. Chen, J.~B. Dent, S.~Dutta, and E.~N. Saridakis, \newblock Class. Quant. Grav. {\bf 28}, 215011 (2011) [arXiv:1104.4349].

\bibitem{Bahamonde_2023}
S. Bahamonde et al., \newblock Rep. Prog. Phys. {\bf 86}, 026901 (2023) [arXiv:2106.13793].

\bibitem{Ashtekar:2006rx}
A.~Ashtekar, T.~Pawlowski, and P.~Singh, \newblock Phys. Rev. Lett. {\bf 96}, 141301 (2006) [arXiv:gr-qc/0602086].

\bibitem{WilsonEwing:2012pu}
E.~Wilson-Ewing, \newblock JCAP {\bf 03}, 026 (2013) [arXiv:1211.6269].

\bibitem{hh}
J.B. Hartle and S.W. Hawkng, Phys, Rev. D {\bf 28}, 2960-2975 (1983).

\bibitem{hertog}
J. Hartle and T. Hertog, (2012) [arXiv:1104.1733].

\bibitem{cc}
B. J. Carr and A. A. Coley, Int. J. Mod. Phys. D. {\bf 20}, 2733-2738 (2011); B. Carr, A. Coley and T. Clifton, (2017) [arXiv:1704.02919].

\bibitem{ccc}
T. Clifton, B. Carr and A. Coley, Class. Quant. Grav. {\bf 34}, 135005 (2017); A. Coley, Class. Quant. Grav. 37, 245002 (2021) [arXiv:2012.14049].

\bibitem{carr-silk}
B. Carr and J. Silk, Mon. Not. Roy. Astron. Soc. \textbf{478}, 3756-3775 (2018); M. Kawasaki and N. Kevlishvili, Nucl. Phys. B {\bf 807}, 229 (2009); A. D. Dolgov, (2016) [arXiv:1605.06749].

\bibitem{cks}
B.~Carr, F.~Kuhnel and M.~Sandstad, Phys. Rev. D \textbf{94}, 083504 (2016) [arXiv:1607.06077].

\bibitem{rovelli}
C. Rovelli and F, Vidotto, (2018) [arXiv:1805.03224].

\bibitem{cainew}
Y.~F.~Cai, C.~Tang, G.~Mo, S.~F.~Yan, C.~Chen, X.~H.~Ma, B.~Wang, W.~Luo, D.~A.~Easson and A.~Marciano, Sci. China Phys. Mech. Astron. \textbf{67}, 259512 (2024) [arXiv:2301.09403].

\bibitem{1}
A. A. Coley and G. F. R. Ellis, Class. Quant. Grav. {\bf 37}, 013001 (2020) [arXiv:1909.05346].

\bibitem{13}
C. A. Mason, M. Trenti and T. Treu, 2023, Mon. Not. Roy. Astr. Soc. {\bf 21}, 497 (2022) [arXiv:2207.14808]; L. Y. A. Yung et al., (2023) [arXiv:2304.04348].

\bibitem{14}
S. L. Finkelstein et al., Astrophys. J. Lett. {\bf 946}, L13 (2023) [arXiv:2211.05792]; Y. Harikane et al., Astrophys. J. Suppl. Ser. {\bf 265}, 5 (2023) [arXiv:2208.01612]; P. Santini et al., Astrophys. J. Lett. {\bf 942}, L27 (2023)[arXiv:2207.11379]; H. Yan et al., Astro phys. J. Letters {\bf 942}, L9 (2023); M. J. Rieke et al., Pub. Astronom. Soc. Pacific {\bf 135}, 028001 (2023)[arXiv:2212.12069].

\bibitem{17}
Labbé, I., van Dokkum, P.G., Nelson, E.J., Bezanson, R.S., Suess, K.A., Leja, J., Brammer, G.B., Whitaker, K.E., Mathews, E.P., Stefanon, M., and Wang, B., Nature {\bf 616}, 266-269 (2022) [arXiv: 2207.12446].

\bibitem{19}
A. Bogdan et al., (2023) [arXiv:2305.15458].

\bibitem{20}
M. Boylan-Kolchin, Nature Astronomy {\bf 7}, 731 (2023) [arXiv:2208.01611]; M. Biagetti, G. Franciolin and A. Riotto, Astrophys. J. {\bf 944}, 113 (2023) [arXiv: 2210.04812].

\bibitem{23}
R.~Sharma and M.~Sharma, Mon. Not. Roy. Astron. Soc. \textbf{531}, 3287-3296 (2024) [arXiv:2310.06898].

\bibitem{corman}
M.~Corman, W.~E.~East and J.~L.~Ripley,
%``Evolution of black holes through a nonsingular cosmological bounce,''
JCAP \textbf{09}, 063 (2022)
[arXiv:2206.08466].

\bibitem{perez}
D.~P{\'e}rez and G.~E.~Romero, Phys. Rev. D \textbf{105}, 104047 (2022), [arXiv:2205.10333].

\bibitem{Mcvittieoriginal}
G.~C. McVittie, \newblock Mon. Not. Roy. Astron. Soc. {\bf 93}, 325 (1933).

\bibitem{A107}
B. C. Nolan, Phys. Rev. D {\bf 58}, 064006 (1998) [arXiv:gr-qc/9805041]; {\em{ibid.}} Class. Quant. Grav. {\bf 16}, 1227 (1999) ; {\em{ibid.}} Class. Quant. Grav. {\bf 16}, 3183 (1999) [arXiv:gr-qc/9907018]; {\em{ibid.}}. Class. Quant. Grav., {\bf 34}, 225002 (2017); {\em{ibid.}}, Class. Quant. Grav. {\bf 16} (1999) ; {\em{ibid.}}, Class. Quant. Grav. {\bf 42}, 235019 (2025) [arXiv:2509.18903].

\bibitem{A110}
Z.-H. Li and A. Wang, Mod. Phys. Lett. A {\bf 22}, 1663 (2007) [arXiv:astro-ph/0607554]; N. Kaloper, M. Kleban and D. Martin, Phys. Rev. D {\bf 81}, 104044 (2010) [arXiv:1003.4777]; R.A. Sussman, Gen. Relativ. Gravit. {\bf 17}, 251 (1985) ; V. Faraoni and M. Rinaldi, Phys. Rev. D {\bf 110}, 063553 (2024) [arXiv:2407.14549]; L. Modesto and E. Rattu, (2025) [arXiv:2510.04165].

\bibitem{Peter_2007}
P.~Peter, E.~J.~C. Pinho, and N.~Pinto-Neto, \newblock Phys. Rev. D {\bf 75} (2007) [arXiv:hep-th/0610205].

\bibitem{Pinto_Neto_2013}
N.~Pinto-Neto and J.~C. Fabris, \newblock Class. Quant. Grav. {\bf 30}, 143001 (2013) [arXiv:1306.0820].

\bibitem{Faraoni_2007}
V.~Faraoni and A.~Jacques, \newblock Phys. Rev. D {\bf 76} (2007) [arXiv:0707.1350].

\bibitem{Nolan_1999}
B.~C. Nolan, \newblock Class. Quant. Grav. {\bf 16}, 3183 (1999) [arXiv:gr-qc/9907018].

\bibitem{Kaloper_2010}
N.~Kaloper, M.~Kleban, and D.~Martin, \newblock Phys. Rev. D {\bf 81} (2010) [arXiv:1003.4777].

\bibitem{Nandra_2012}
R.~Nandra, A.~N. Lasenby, and M.~P. Hobson, \newblock Mon. Not. Roy. Astron. Soc. {\bf 422}, 2931 (2012).

\bibitem{Gregoris_2020}
D.~Gregoris, Y.~C. Ong, and B.~Wang, \newblock Eur. Phys. J. C {\bf 80} (2020).

\bibitem{faraoni_2018}
V.~Faraoni, (2018) [arXiv:1810.04667].

\bibitem{P_rez_2021}
D.~P{\'e}rez, S.~E.~P. Bergliaffa, and G.~E. Romero, \newblock Phys. Rev. D {\bf 103} (2021) [arXiv:2103.00108].

\bibitem{Valentino}
E. Di Valentino, et al, Class. Quant. Grav. {\bf 38}, 153001 (2021) [arXiv:2103.01183]; E. Di Valentino et al., Phys. Dark Univ. {\bf 49}, 101965 (2025) [arXiv:2504.01669]; S Verma, A Dixit, A Pradhan, M. S. Barak JHEP {\bf 49}, 100440 (2026) [arXiv:2508.20107].

\bibitem{TorBounce}
Y.-F. Cai, S.-H. Chen, J.B. Dent, S. Dutta and E.N. Saridakis, Class. Quant. Grav. {\bf 28} (2011) 215011 [arXiv:1104.4349]; S. K. Tripathy, Sasmita Pal, B. Mish, Eur. Phys. J. C {\bf 84}, 1202 (2024) [arXiv:2410.23307]; S. Alam, S. Sen J. M. Lamia, and S. Sengupta, (2025) [arXiv:2509.03508].

\bibitem{aldrovandi1732013}
R.~Aldrovandi and J.~G.~Pereira, Teleparallel Gravity: An Introduction,
Springer, (2013).

\bibitem{krssak362019}
M. Krssak et al., Class. Quantum Grav. {\bf 36}, 183001 (2019)[arXiv:1810.12932].

\bibitem{coley616439}
A. A. Coley, R. J. van den Hoogen, and D. D. McNutt, J. Math. Phys. {\bf 61}, 072503 (2020) [arXiv:1911.03893]; {\bf 64}, 032503 (2023) [arXiv:2302.11493]; Class. Quantum Grav. {\bf 39}, 22LT01 (2022) [arXiv:2205.10719]; A. A. Coley et al., Eur. Phys. J. C {\bf 84}, 334 (2024) [arXiv:2402.07238].

\bibitem{diego}
D.~F.~L{\'o}pez, A.~A.~Coley and R.~J.~van den Hoogen, (2025) [arXiv:2508.20314].

\bibitem{coleypersistance}
B.~J. Carr and A.~A. Coley, \newblock Int. J. Mod. Phys. D {\bf 20}, 2733 (2011) [arXiv:1104.3796].

\bibitem{McNutt:2023nxm}
D.~D.~McNutt, A.~A.~Coley and R.~J.~v.~Hoogen, J. Math. Phys. \textbf{64}, 032503 (2023)
[arXiv:2302.11493].

\end{thebibliography}
\end{document}